\documentclass[superscriptaddress,twocolumn,prb]{revtex4}

\usepackage{graphicx}
\usepackage{subfigure}

\usepackage{amsmath}
\usepackage{amssymb}
\usepackage{amsthm}
\usepackage{ulem} 

\begin{document}

\title{Predictive GW calculations using plane waves and pseudopotentials}

\author{J. Klime\v{s}}
\affiliation{University of Vienna, Faculty of Physics and Center for 
Computational Materials Science,
 Sensengasse 8/12, A-1090 Vienna, Austria}

\author{M. Kaltak}
\affiliation{University of Vienna, Faculty of Physics and Center for
Computational Materials Science,
 Sensengasse 8/12, A-1090 Vienna, Austria}

\author{G.~Kresse}
\affiliation{University of Vienna, Faculty of Physics and Center for 
Computational Materials Science,
 Sensengasse 8/12, A-1090 Vienna, Austria}

\begin{abstract} 

We show that quasiparticle (QP) energies  as calculated
in the $GW$ approximation converge to the wrong value
using the projector augmented wave (PAW) method, 
since the 
overlap integrals between occupied orbitals and high energy,
plane wave like orbitals, are incorrectly described. 
The error is shown to be related to the incompleteness of
the partial wave basis set inside the atomic spheres. 
It can  be avoided by adopting norm-conserving partial waves,
as shown by analytic expressions for
the contribution from unoccupied orbitals with high kinetic energy. 
Furthermore,  $G_0W_0$ results based on norm-conserving PAW potentials 
are presented for a large set of semiconductors and insulators.
Accurate extrapolation procedures to the infinite basis set limit and infinite
k-point limit are discussed in detail.
\end{abstract}

\maketitle

\section{Introduction}

Density functional theory (DFT) does not allow to calculate
accurate quasiparticle (QP) energies as measured experimentally
in X-ray photoelectron spectroscopy (XPS) or ultraviolet photoelectron spectroscopy (UPS),
although the Kohn-Sham eigenvalues often show at least qualitative
agreement with the corresponding measurements.
In solid state physics, the standard approach to determine
QP energies is currently the $GW$ approximation often combined with 
orbitals from a Kohn-Sham DFT calculation,
for instance the local density approximation (LDA).
The $GW$ method was originally proposed by Hedin\cite{hedin1965} and
later extensively applied by the group of Louie,\cite{hybertsen1986} following
a procedure first adopted by Sham and Schl\"uter for
tight binding models.\cite{lannoo1985} For recent reviews we
refer to Refs.~\onlinecite{aryasetiawan1998,onida2002,giuliani2005,bechstedt2009},
and for a recent re-derivation we refer to Ref.~\onlinecite{starke2012}.
With the advances in compute power over the last two decades,
the $GW$ method can now be routinely applied to systems containing
hundred atoms, and with some approximations, calculations for several hundred
atoms are possible or will be possible in the near future.

Remarkably, even after several decades of research, 
publications with technically converged QP energies are still rare and seem
to pose a significant challenge to modelers. In part, this is
related to the slow convergence with respect to the number
of k-points used to sample the Brillouin zone. However, more severe
is the slow convergence of the QP energies with respect to the basis
set size. This problem has gone unnoticed for some time, with 
most researchers using a few hundred unoccupied orbitals
per atom in the calculation of the Green's function. 
However, the recent work of Shih~{\it et al.} on ZnO suggests 
that convergence can be exceedingly slow requiring thousands of orbitals.\cite{shih2010} 
Furthermore, care must be taken to reach convergence with respect
to both the plane wave basis set for the orbitals and
the plane wave basis set for the response function.\cite{shih2010}
Although the ZnO results
were somewhat exaggerating the dependence on the number of orbitals and 
the final values for the band gap of ZnO were not very accurate--- partly because
of the use of a simplified plasmon-pole model\cite{stankovski2011} ---the calculations
still mark a turning point: whenever highly accurate results
are required for the QP energies, huge basis sets are
needed to obtain them.

The situation is very similar to quantum chemistry (QC) calculations,
where convergence of the total energies (and excitation energies)
is also exceedingly slow with the basis set size. 
In the QC community, it is well established that the inter-electron 
cusp--- the dependence of the many electron wave function 
on the distance of two electrons 
---causes slow convergence.\cite{kato1957,schwartz1962,kutzelnigg1992,kendall1992}
As a consequence of a wavefunction kink at zero distance, absolute correlation energies 
show convergence proportional to the inverse of the number of basis functions.\cite{halkier1998}
A similar slow convergence was observed for the correlation energy
in the random phase approximation (RPA). 
Based on calculations using the Lindhard dielectric
function but without explicit derivations,  Harl and
Kresse suggested that the RPA correlation energy converges like
$1/N$, where $N$ is the plane wave basis set size.\cite{harl2008} 
For MP2 total energies, a similar convergence behavior  was derived by
Shepherd~{\it et al.}, closely paralleling
the finding for quantum chemistry basis sets.\cite{shepherd2012}
Gulans explicitly derived the same convergence behavior for the RPA correlation energy
relying on the jellium model
and showed that a similar $1/N$ convergence 
must be expected for the QP energies in solids.\cite{gulans_unp,bjorkman2012}
Although his derivation suggests that the origin of the slow convergence 
of QP energies and total correlation energies is related,  
how exactly the cusp-condition appears in methods based on
one-electron Green's functions is somewhat ``clouded". 
In fact, in one-electron Green's functions approaches, the self-energy needs to
account for the correlation hole or, equivalently, the screening charge, 
which relates at short distances to the cusp condition.
Another way to understand this is that QP energies essentially represent
the difference between two total energy calculations, each with a $1/N$ convergence.

In this work, we derive a closed expression for the MP2 energy
and self-energy in the limit of high energy states (long plane wave vectors). 
The derivation does not rely on an  explicit form for the dielectric function 
for the jellium electron gas.
We only assume that for high energies, well above the vacuum level,
the one electron states become essentially plane waves. The correction turns
out to depend on the charge density distribution alone.
Interestingly, the present derivation also
shows that the Coulomb-hole plus screened-exchange (COHSEX) 
approximation overestimates the high energy contributions to the QP energies by
a factor two, as observed recently by Kang and Hybertsen.\cite{kang2010}
Furthermore, in order to obtain converged values, one needs to increase 
both the basis set for the orbitals, as well as the 
basis set for the response function.
Increasing only one of them leads to a incorrect limiting behavior.

Unfortunately, the present derivation also highlights a severe problem
for ultrasoft pseudopotentials and the related projector augmented wave potentials that fail to conserve the
norm in the plane wave representation. 
For groundstate properties, violation of the norm 
is usually eliminated by adding appropriate augmentation or compensation charges
centered at each atomic site. 
However, such a correction is missing  for plane wave like orbitals 
at very high kinetic energies. 
As a result, the overlap integrals between ground state orbitals
and unoccupied orbitals at very high kinetic energies are not properly described. 
A simple solution to this problem is to construct potentials that are norm-conserving. 
In the present work, this is achieved by determining projector augmented wave
potentials with norm-conserving partial waves. 
We present tests for ZnO, GaAs, and AlAs that indicate that QP energies 
extrapolated to the infinite basis set limit are then independent of 
the detailed construction of the potential. 
The final gaps agree very well with previous calculations 
of Friedrich {\it et al.} for ZnO.\cite{friedrich2011}
Motivated by this result, we apply the present procedure to 24 semiconductors 
and insulators and present $G_0W_0$ as well as $GW_0$  calculations based 
on LDA and PBE (Perdew Burke Ernzerhof)\cite{perdew1996} orbitals. 
We believe that these results are technically converged and can 
serve as a stringent benchmark for future $GW$ calculations and 
validation of different implementations.

\section{Theory}

In the following sections we will derive the 
contributions to the total many electron correlation energy and self-energy from
wave vectors  at very high kinetic energies.
The important assumption is that the orbitals at sufficiently high energies
are well approximated by plane waves. 
This is, however, a good approximation, since at sufficiently high energies, 
the kinetic energy term $-\hbar^2 \Delta/(2m) $ will be the dominant part
of the Hamiltonian so that the orbitals become ultimately plane waves. 
Our derivation first concentrates on the total correlation energies in the
second order approximation (direct term in MP2) since
the essential results are easy to grasp for this case. Then we discuss
the shortcomings of the PAW method, and finally move on
to the self-energy.

\subsection{Correlation energy in second order
perturbation theory (MP2)}

The direct contribution to the MP2 correlation energy can be written as
\begin{equation}
\label{equ:MP2}
\begin{split}
E^{\rm dMP2} & = -2\sum_{ij}^{\rm occ} \sum_{ab}^{\rm unocc} \sum_{GG'} \langle i | G | a \rangle \langle a | -G' | i \rangle {4\pi e^2\over \Omega G'^2} \times \\
 & \langle j | G' | b \rangle \langle b | -G | j \rangle {4\pi e^2\over \Omega G^2} {1\over \varepsilon_a +  \varepsilon_b -  \varepsilon_i -  \varepsilon_j}\,.
\end{split}
\end{equation}
Here, $G$  and $G'$ are three dimensional vectors,
and the spin coordinates have been summed over so that
the summation over orbitals is over spatial orbitals only.
The integers $i, j$ and $a, b$ are indices for occupied and unoccupied orbitals respectively, and 
\begin{equation}
\label{equ:overlapdens}
\langle b | -G | j \rangle=\int_\Omega \psi_b^*(r) \psi_j(r) \exp(-iGr) \, {\rm d}r
\end{equation}
is the overlap (transition) density of the orbitals
$ \langle r |b \rangle = \psi_b(r)$ and $ \langle r |j \rangle = \psi_j(r)$
in a basis of plane waves, with $r$ being a three dimensional vector.

Our derivation is for the direct MP2 (dMP2) energy, however, in the limit of long wave vectors,
the RPA correlation energy also reduces to this form,
since screening is rather ineffective at long wave vectors. 
Thus the RPA correlation (here $\chi_0$ and $v$ are the independent particle 
polarizability and the Coulomb operator respectively)
\[
 E^{\rm RPA} = Tr[ \frac{1}{2} (\chi_0 v)^2 +  \frac{1}{3} (\chi_0 v)^3 + ...] \approx
\frac{1}{2} (\chi_0 v)^2
\]
reduces to the direct MP2 form, since $\chi_0$ becomes small for large $G$ vectors. 
The relation between the direct MP2 and
RPA energy has also been discussed in Ref.~\onlinecite{marsman2009}.

We now determine the contribution $E^{\rm dMP2}(G',G)$ to 
the direct MP2 energy from plane wave vectors
$G$ and $G'$ that have {\rm no overlap with occupied ground state orbitals}.
For brevity, the superscript ${\rm dMP2}$ will be dropped from now on.
Specifically, we want to calculate the approximate correlation energy
from plane waves $G$ and  $G'$ that observe
\begin{equation}
\label{equ:assumption}
\langle i |  -G | j \rangle \approx 0 
\end{equation}
for all occupied orbitals $i,j$. 
Later for QP calculations, we will extend this condition to few unoccupied states of interest. 
These plane wave components are obviously not relevant for
the description of the occupied Kohn-Sham orbitals (or their overlap density), 
but as we show below they still contribute to the many body correlation energy and 
self-energy.

In a first step, we allow the indices $a$ and $b$ to run over all
states instead of the unoccupied states  in Eq.~(\ref{equ:MP2}) 
(Eq.~(\ref{equ:assumption}) allows for this ``simplification"):
\begin{equation}
\label{equ:MP2G}
\begin{split}
&E(G',G)  = -2\sum_{ij}^{\rm occ} \sum_{ab}^{\rm all} \langle i | G | a \rangle \langle a | -G' | i \rangle {4\pi e^2\over\Omega G'^2}  \times \\
 & \langle j | G' | b \rangle \langle b | -G | j \rangle {4\pi e^2\over \Omega G^2} {1\over \varepsilon_a +  \varepsilon_b -  \varepsilon_i -  \varepsilon_j}\,.
\end{split}
\end{equation}
Second, we assume  that high energy unoccupied states $a$ are 
essentially plane waves of the form $\psi_a(r) ={1\over \sqrt{\Omega}} \exp(iG_ar)$ with
the eigen energies $\varepsilon_a=(\hbar G_a)^2/(2m)$.
In this case, the quantities $\langle i | G | a \rangle \langle a | -G' | i \rangle$ and $\langle j | G' | b \rangle \langle b | -G | j \rangle $ are only sizable if $G_a \approx -G$ and
$G_b \approx -G'$, or $G_a \approx G'$ and $G_b \approx G$ since the occupied states have predominantly components at small wave vectors.
Therefore, we can approximate $ \varepsilon_a +  \varepsilon_b \approx \hbar^2(G^2 + G'^2)/(2m)$.
Moreover, we can neglect the dependence on the energies of occupied states compared to those of the unoccupied states, so that
\[ 
\varepsilon_a +  \varepsilon_b -  \varepsilon_i -  \varepsilon_j \approx \hbar^2(G^2 + G'^2)/(2m)\,.
\]
This approximation is also correct for the important diagonal components $G = G'$.
Once rid of the eigenvalues $\varepsilon$, the resolution of identity can be
used  $\sum_a  | a \rangle \langle a | = 1$ leading to a much simpler expression
\begin{equation}
\begin{split}
E(G,G')= -{2m \over \hbar^2}\sum_{ij}^{\rm occ} \langle i | G -G' | i \rangle {4\pi e^2\over \Omega G'^2}\times \\
\langle j | G' -G | j \rangle {4\pi e^2\over \Omega G^2} {2\over G^2+G'^2}\,.
\end{split}
\end{equation}
We note again that this is an approximation, but it becomes exact, if the 
plane waves $G$ and $G'$ are limited to sufficiently long wave vectors.
We can also write this as 
\begin{equation}
\begin{split}
E(G,G')=  -{2m \over \hbar^2} \sum_{ij}^{\rm occ} \rho_i(G -G')  {4\pi e^2\over G'^2} \times\\ \rho_j(G' -G)   {4\pi e^2\over G^2} {2\over G^2+G'^2}\,,
\end{split}
\end{equation}
where the Fourier transformed density of the orbital  $i$ is defined as:
\begin{equation}
\rho_i(G) = {1\over \Omega} \int_\Omega \psi_i^*(r) \psi_i(r) \exp(i Gr) \,{\rm d}r \; .
\end{equation}
Since the total density can be calculated as $ \rho(G) =2\sum_i \rho_i(G)$, we obtain the final expression
\begin{equation}
\label{eq_dmp2_rho2}
E(G,G')= - {m\over 2\hbar^2}|\rho (G -G')|^2  {4\pi e^2\over G'^2}  {4\pi e^2\over G^2}  {2\over G^2+G'^2}\,.
\end{equation}
For the jellium electron gas with $\rho (G -G')= \delta(G -G') \rho(0)$ this
simplifies to 
\begin{equation}
\label{eq_dmp2_cc}
E (G,G')= - {m\over 2\hbar^2}\delta(G,G') \rho (0)^2  {4\pi e^2\over G^2}  {4\pi e^2\over G^2} {1\over G^2}\,.
\end{equation}
where $\rho (0)$ is the electron density.
From this derivation, it is clear that, in principle, the groundstate density distribution 
and its Fourier transform alone determine the asymptotic contribution to the correlation energy.
Summing over all plane waves  for  $|G|,|G'|> G_{\rm cut} $ is trivial and yields
\begin{equation}
\Delta E= \sum_{|G|>G_{\rm cut}} \sum_{|G'|>G_{\rm cut}}  E (G,G') \approx {4 m e^4\over3 \hbar^2} {\rho^2\over G_{\rm cut}^3}\,.
\end{equation}
This can be also written using the number of plane waves $N_{\rm pw}$ in the cut-off sphere with radius $G_{\rm cut}$
\begin{equation}
\label{eq_dmp2_ccfin}
\Delta E
\approx {2 m e^4\over 9 \pi^2 \hbar^2} {\rho^2 \Omega^2\over N_{\rm pw}}
={2 m e^4\over 9 \pi^2 \hbar^2} {N_{\rm el}^2\over N_{\rm pw}},
\end{equation}
where $N_{\rm el}$ is the total number of electrons in the cell (including spin).
Therefore, the error decays asymptotically with the inverse of $N_{\rm pw}$.
The expression can be obtained for a non-uniform electron distribution as well in which case
is $\rho^2$ replaced by $\sum_g |\rho(g)|^2$, where $g=G-G'$.
To derive this we assumed that the groundstate density has only contributions at
wave vectors $g$ that are much smaller than wave vectors $G$ and $G'$.
The same dependence of the error on $N_{\rm pw}$ was also derived in Ref.~\onlinecite{shepherd2012}. 
It is important to note that the slow convergence of the MP2 energies is
well established for quantum chemistry basis sets (Gaussian type orbitals) with exactly
the same one over basis set convergence.\cite{halkier1998}
Furthermore, it has  been shown that the slow convergence is related 
to the description of the two electron cusp condition.\cite{kato1957,schwartz1962,kutzelnigg1992,kendall1992,gulans_unp}
This condition leads to a kink in the many electron wave function $\Psi(r_1,r_2, r_3, ...)$:
when any two coordinates $r_i$ and $r_j$ coincide, the
many electron wavefunction exhibits a discontinuity in the slope, which leads  
exactly to the basis set convergence determined above. 

Interestingly, the convergence of the correlation energy is entirely unrelated to the  basis set size required to  
calculate the groundstate one electron orbitals (compare Eq.~\ref{eq_dmp2_rho2}).
Instead, the slow convergence originates from the  fact that in the correlation energy the density
enters at wave vectors $g=G-G'$. 
This implies that long wave vectors are folded back to wave vectors around $g=0$. 
Hence, even very long wave vectors not relevant for groundstate calculations will contribute
to the final many electron correlation energy.  
Since the correlation energy contribution is determined by the charge density, one can also not simply
claim that this energy contributions will cause a trivial shift of the total correlation energy. 
Obviously, when the density changes, the convergence rate will be altered as well.

Looking back at the derivation a second related issue becomes obvious.
We expect to see the slow convergence only, if sufficiently
long wave vectors are included in both basis sets, for the orbitals
$\psi_a(r) ={1\over \sqrt{\Omega}} \exp(iG_ar)$, with $G_a \approx G$,
as well as for the overlap densities  $\langle i | G | a \rangle$.
If the basis for the orbitals is truncated at $G_{\rm cut}^{pw}$, or
if the basis for the overlap densities is truncated at $G_{\rm cut}^{\rm \chi}$,
an error proportional to $1/ G_{\rm cut}^3$ is introduced with $G_{\rm cut} = \min(G_{\rm cut}^{pw}, G_{\rm cut}^{\rm \chi})$.
It is therefore not meaningful to improve only the one-electron
basis set or the basis set for the overlap densities alone
as this leads to a ``false" limiting behavior.

We also note that Eq.~(\ref{eq_dmp2_ccfin}) gives the leading
order of the error which behaves like $1/N_{\rm pw}$.
As discussed by Gulans, the next order correction falls off like $N_{\rm pw}^{-5/3}$.\cite{gulans_unp}
This order would be recovered if the energies of the occupied states were taken into account
or, in the general case, by not assuming that $G^2=G'^2$ in the Coulomb and energy terms of Eq.~(\ref{eq_dmp2_rho2}).
The next order follows $N_{\rm pw}^{-7/3}$ and such dependence of MP2 energy was found by Gr\"{u}neis~{\it et al.}
after applying F12 corrections.\cite{gruneis2013} 
Presumably, the F12 corrections treat the error of the order $N_{\rm pw}^{-5/3}$ exactly
by including both the energies and basis set representation of the occupied states. 
Another contribution of the order $N_{\rm pw}^{-7/3}$ comes from the third order interaction,
see Ref.~\onlinecite{gulans_unp}, this appears in the RPA calculations but not in the dMP2
correlation energies.
In principle, one can improve the accuracy of the scheme by keeping the energies of the occupied states 
in the formulae, we leave, however, the explicit derivation for future work.

\subsection{Shortcomings of the PAW method: Completeness issues }

In the PAW method the all electron orbitals $|i\rangle$ are related to the pseudo part $|\tilde i\rangle$ by a linear relation
\begin{equation}
\label{equ:PAW}
|i\rangle=|\tilde i\rangle+ \underbrace{\sum_{\alpha} (|\alpha\rangle-|\tilde \alpha\rangle)\langle p_\alpha|\tilde i\rangle}_{| i^{\rm aug}\rangle} \,,
\end{equation}
where $\langle p_\alpha|$ is an atom centered projector onto an atomic orbital  $\alpha$ with corresponding all-electron and pseudo partial waves 
$\langle r | \alpha \rangle = \phi_\alpha(r)$ and $\langle r | \tilde \alpha \rangle =  \tilde \phi_\alpha(r)$, respectively.
The crucial approximation in the PAW method is that the set of 
partial waves is complete. 
This allows to simplify one electron quantities such
as the overlap density significantly: obviously, any one electron 
expectation value  such as $\langle a | r | i \rangle$ would in principle involve 
a term $\langle \tilde a | r | \tilde i \rangle$, one with the augmentation terms
$ \langle a^{\rm aug} | r| i^{\rm aug}\rangle$, and mixed terms. 
Completeness allows to drop the mixed terms as discussed in Ref.~\onlinecite{bloechl:94}. 
This simplifies the overlap density to
\begin{equation}
\label{equ:aug1}
\langle a | r | i \rangle \approx \langle \tilde a | r | \tilde i \rangle
+ \sum_{\alpha\beta} \langle \tilde a|p_{\alpha}\rangle
(\langle \alpha| r |\beta\rangle-\langle \tilde \alpha| r |\tilde \beta\rangle)\langle p_\beta|\tilde i\rangle,
\end{equation}
which is crucial for the efficiency of the PAW method. 
It is easy, within the PAW method, to achieve completeness for ground state orbitals, however, 
it is in practice impossible to achieve completeness for states at very high energies.
This is the source of an important error.

Returning to Eq.~(\ref{equ:MP2G}), the problem of the PAW method is  easy to understand. 
If $a$ is a plane wave with high kinetic energy $\psi_a(r) ={1\over \sqrt{\Omega}} \exp(iG_ar)$,  
the projection $\langle a | p_\alpha \rangle$ becomes very small, since the projectors $| p_\alpha \rangle$
span only the groundstate orbitals and are limited in Fourier space ($\langle  G_a| p_\alpha \rangle \approx 0$). 
The completeness relation is then violated, and, in principle, the mixed 
terms discussed above should be reintroduced to alleviate  the issue. 
As long as the mixed terms are not included, the augmentation terms are
effectively zero between groundstate orbitals and unoccupied plane wave like 
orbitals at high energies, so that 
$\langle i | G | a \rangle \langle a | -G' | i \rangle$ becomes identical
to $\langle \tilde i | G | \tilde a \rangle \langle \tilde a | -G' | \tilde i \rangle$.
In this case, the correlation energy in the limit of long wave vectors $G$ and $G'$
becomes
\begin{equation}
E(G,G')= - {me^4\over \hbar^2} |\tilde \rho (G -G')|^2  {4\pi\over G'^2}  {4\pi\over G^2} {1\over G^2+G'^2}\,,
\end{equation}
where $\tilde \rho (G -G')$ is the groundstate density neglecting any
charge augmentation. 
This can result in a sizable error, since for  $d$ elements the norm 
violation is often significant, up to  80~\% for  late $3d$ elements 
such as Zn, Ga, or Cu.
While one still observes an error decreasing as $1/N_{\rm pw}$, the limiting 
value is incorrect.

\subsection{Second order self-energy contribution at large $G$ vectors}

The usual definition for the self-energy in the $GW$ approximation is
\begin{equation}
\label{eq_sigma}
\Sigma(G,G',\omega)={i\over 2\pi} \int {\rm d} \omega' \exp(-i\delta\omega') G(\omega-\omega') W(\omega')\,,
\end{equation}
with the Green's function
\begin{equation}
G(G,G',\omega)=\sum_n^{\rm all} {\langle G | n \rangle \langle n | G' \rangle \over \omega - \varepsilon_n \mp i \eta}\,,
\end{equation}
where $\eta$ is a positive infinitesimal, and the minus sign applies to occupied states 
and the plus sign to the unoccupied ones.
As before, we use the second order approximation, which gives for the screened interaction
$W= v \chi_0 v$ (the exchange contribution is not considered here). 
This yields for $W(G,G',\omega)$ (from now on we will 
drop the functional parameters $G, G'$, and $\omega$)
\begin{equation}
\begin{split}
W  =& {2\over \Omega}{4\pi e^2\over G^2}{4\pi e^2\over G'^2}  \sum_i^{\rm occ}\sum_a^{\rm unocc} 
 \Big( {{\langle i | -G | a \rangle \langle a | G' | i \rangle} \over {\omega + \varepsilon_i - \varepsilon_a + i \eta}}  \\
& -{{\langle a | -G | i \rangle \langle i | G' | a \rangle} \over { \omega + \varepsilon_a - \varepsilon_i - i \eta} }  \Big) \,.
\end{split}
\end{equation}
Inserting these expressions into Eq.~(\ref{eq_sigma}) one obtains 
\begin{multline}
\label{eq_sigma2}
\Sigma={i\over 2\pi} \int {\rm d} \omega' \exp(-i\delta\omega') 
\sum_n {\langle G | n \rangle \langle n | G' \rangle \over \omega -\omega'- \varepsilon_n \mp i \eta}\times
\\
{2\over \Omega}{4\pi e^2\over G^2}{4\pi e^2\over G'^2}  \sum_{i,a} 
\left( {{\langle i | -G | a \rangle \langle a | G' | i \rangle} \over {\omega' + \varepsilon_i - \varepsilon_a + i \eta}}
-{{\langle a | -G | i \rangle \langle i | G' | a \rangle} \over { \omega' + \varepsilon_a - \varepsilon_i - i \eta} }  \right).
\end{multline}
The self-energy integral can now be 
calculated analytically via contour integration considering the poles of the integrand.
As we choose to close the contour in the lower half, only the poles with a negative imaginary part will contribute.
There are two distinct contributions, the first originates from the
poles in the Green's function at the energies of the occupied states
\begin{equation}
\omega'=\omega-\varepsilon_n-i\eta
\end{equation}
that give the screened exchange contribution (we drop $\eta$ from the expressions after integration)
\begin{multline}
\label{equ:SEX}
\Sigma^{\rm SEX(2)}= -{2\over \Omega}{4\pi e^2 \over G^2} {4\pi e^2\over G'^2}  \sum_n^{\rm occ} {\langle G | n \rangle \langle n | G' \rangle}  \times
 \\
\sum_{i,a}
\left( {{\langle i | -G | a \rangle \langle a | G' | i \rangle} \over {\omega -\varepsilon_n + \varepsilon_i - \varepsilon_a}}
-{{\langle a | -G | i \rangle \langle i | G' | a \rangle} \over { \omega -\varepsilon_n + \varepsilon_a - \varepsilon_i } }  \right) \,.
\end{multline}
The second contribution comes from the poles of 
$W$ located at the excitation energies of the non-interacting response function
\begin{equation}
\omega'=-(\varepsilon_i - \varepsilon_a + i \eta)\,.
\end{equation}
This gives the Coulomb hole contribution  to the self-energy
\begin{multline}
\label{equ:COH}
\Sigma^{\rm COH(2)}=  {2\over \Omega}{4\pi e^2\over G^2} {4\pi e^2\over G'^2} \sum_n^{\rm all} {\langle G | n \rangle \langle n | G' \rangle }\times
 \\
\sum_{i,a} {\langle i | -G | a \rangle \langle a | G' | i \rangle \over \omega+\varepsilon_i-\varepsilon_a - \varepsilon_n } \,. 
\end{multline}
These expressions are equivalent to the ones derived by Gr\"{u}neis~{\it et al.}\cite{gruneis2010}
However, we have kept the terms with zero imaginary energy in the COH and SEX parts,
(i.e., the first term in the SEX and the part where $b\in{\rm occ}$ in the COH). 
These are identical differing only in the sign and will subtract to zero
in the COHSEX approximation.

We now consider the contribution of the long wave vectors $G$ and $G'$. 
When the SEX term  is evaluated for an orbital $m$ with eigen energy $\varepsilon_m$, 
$\langle m | \Sigma^{\rm SEX(2)} | m \rangle$, 
the first term on the r.h.s. of Eq.~(\ref{equ:SEX}) contains
\[
 \sum_n^{\rm occ} \langle m|  G | n \rangle \langle n | -G' | m \rangle.
\]
Here, $n$ corresponds to an occupied state, and  $m$ corresponds to an occupied state or 
one just above the Fermi-level.
According to our assumption  Eq.~(\ref{equ:assumption}),  at large
$G$ and $G'$ the corresponding contribution is zero. The SEX
term, therefore, does converge rapidly with respect to the basis set size.

The COH contribution, however, involves a summation over all states,
and the matrix element of self-energy reads 
\begin{equation}
\begin{split}
\label{eq_coh2_eval}
\langle m | \Sigma^{\rm COH(2)}(\varepsilon_m)| m\rangle =  {2\over \Omega}{4\pi e^2\over G^2} {4\pi e^2\over G'^2}\times \\
 \sum_n^{\rm all} {\langle m| G | n \rangle \langle n | G' | m\rangle } 
 \sum_{i,a} {\langle i | -G | a \rangle \langle a | G' | i \rangle \over \varepsilon_m+\varepsilon_i-\varepsilon_a - \varepsilon_n } \,. 
\end{split}
\end{equation}
For large $G$ vectors we can modify this expression in the same way as in the MP2 total energy case,
so that for orbital $m$, the approximate contribution to the QP correction can be written as
\begin{equation}
\label{equ:exactGW}
\Delta \varepsilon_m=- {m e^4\over2 \hbar^2} \rho_m(G -G')  {4\pi\over G'^2} \rho(G' -G)   {4\pi\over G^2} {2\over G^2+G'^2}\,.
\end{equation}
This is the final result, and the formal equivalence to the MP2 total energy contribution is obvious. 
In fact, one could have derived this equation by applying Koopmans' theorem to the total energy expression
Eq.~(\ref{eq_dmp2_rho2}). 
After performing the summation over the omitted $G$ vectors the correction can be written in the same manner as in the MP2 case
\begin{equation}
\label{equ:exactGWN}
\Delta\varepsilon_m=-{2 m e^4\over 9 \pi^2 \hbar^2} {2N_{\rm el}\over N_{\rm pw}}\,.
\end{equation}
Again, $N_{\rm pw}$ can be either the number of basis set functions in the response function or the number
of included bands. 
In the case of a non-uniform density, $2N_{\rm el}$ is replaced by $4\sum_g \rho_m(g) \rho(-g)\Omega^2$, where the summation
over the reciprocal lattice vectors $g$ is performed.

It is interesting to relate the second order quasiparticle equation to the commonly employed static 
COHSEX in second order.
The static COH contribution is obtained by assuming $\omega-\varepsilon_n=0$ (or $\varepsilon_m-\varepsilon_n=0$), 
{\rm i.e.}, assuming that the largest contributions to the self-energy come from states close in energy to the state $m$.
In second order, this yields the following static COH contribution:
\begin{multline}
\langle m | \Sigma^{\rm stat-COH(2)}| m \rangle= {2 m e^4\over \hbar^2 \Omega}{4\pi \over G^2} {4\pi \over G'^2} \times
\\
 \sum_n^{\rm all} {\langle m |-G | n \rangle \langle n | G'| m \rangle }
\sum_{ia} 
{\langle i | -G | a \rangle \langle a | G' | i \rangle \over \varepsilon_i-\varepsilon_a }\,. 
\end{multline}
For large $G$ and $G'$ a sizable contribution is only expected to arise 
if the orbital $a$ corresponds to a plane wave  $G_a= -G$.
Neglecting the energy of the occupied state $\varepsilon_i$, one obtains  $\varepsilon_i-\varepsilon_a \approx \hbar^2G^2/(2m)$,
exactly twice the proper value in the full second order equation.
Hence,  the static COHSEX overestimates the high $G$ contributions by a factor 2, 
as has been recently observed in actual calculations (without proof) by 
Kang and Hybertsen.\cite{kang2010} 
We note, that the factor $1\over2$ can  be also derived using the plasmon pole approximation 
as shown by Deslippe~{\it et al.} following a route similar to our current derivation.\cite{deslippe2013}
The present derivation, however, is exact in second order and does not
involve any specific model (except the assumption that
at high energy, plane waves do not contribute to
the expansion of the groundstate orbitals).

\subsection{Shortcomings of the PAW method: Completeness issues again }

There is very little to add compared to the previous discussion on
the correlation energy. 
For long wave vectors, $G$ and $G'$,
the exact contribution to the QP correction for the  state $m$
is given by Eq.~(\ref{equ:exactGW}).
However, in the PAW method we will observe only the pseudised contribution
\begin{equation}
\Delta \tilde \varepsilon_m=-   {m e^4\over 2\hbar^2} \tilde \rho_m(G -G')  {4\pi\over G'^2} \tilde \rho(G' -G)   {4\pi\over G^2} {2\over G^2+G'^2}\,.
\end{equation}
Contributions from the augmentation charges are again missing
since states at large plane waves are not picked up by the projectors.
In principle, it is possible to correct for this error {\rm a posteriori}, by calculating
the exact term (\ref{equ:exactGW}) and subtracting the
one effectively used in the PAW approximation.
In the present case this correction has not been applied.

Instead, Eq.~(\ref{equ:exactGW}) suggests a simple alternative procedure. 
If the pseudized density  $\tilde \rho(g)$ has
the correct norm, both $ \rho(g)$ and $\tilde \rho(g)$ will
exactly agree at $g=0$, and approximately at larger wave vectors. 
Hence, norm-conservation  should help to get a reasonable approximation for the self-energy. 
In other words, if the norm of the pseudized partial waves and all-electron partial waves are identical, 
we expect  smaller error  than if the norm is not conserved. 
This conjecture will be validated by the tests shown in Sec. \ref{sec:ZnO}.

\section{Computational Details}

\begin{table}[t]
\caption{
PAW potentials used in the present work.
The columns r$_{s,p,d,f}$ specify the core radii
for each angular quantum number in a.u.
The ``default'' plane wave cut-off energy $E_{\rm cut}^{pw}$ for the orbitals is specified in eV.
Column ``local'' specifies the chosen local potential. This is usually
the all-electron potential replaced by a soft approximation inside
the specified core radius or the $f$ potential.
}
\label{tab:S2}
\begin{ruledtabular}
\begin{tabular}{lccccccc}
        &  r$_s$  &   r$_p$  &   r$_d$   &  r$_f$  & local &  $E_{\rm cut}^{pw}$ \\
\hline
B       &  1.10   &  1.10    &   1.10    &   -     &  $d$  &  700    \\ 
C	&  1.00   &  1.10    &   1.10    &   -     &  0.8  &  740    \\ 
N	&  0.90   &  1.10    &   1.10    &   -     &  0.9  &  755    \\ 
O	&  1.00   &  1.10    &   1.10    &   -     &  0.9  &  765    \\ 
Mg      &  1.15   &  1.65    &   1.65    &   -     &  1.2  &  821    \\
Al      &  1.75   &  2.00    &   1.80    &  2.00   &  1.4  &  571    \\
Si 	&  1.70   &  1.95    &   1.70    &  2.00   &  1.4  &  609    \\ 
P	&  1.70   &  1.95    &   1.70    &  2.00   &  1.4  &  616    \\ 
S       &  1.23   &  1.35    &   1.70    &  1.80   &  $f$  &  486    \\ 
Zn  	&   1.27  &  1.70    &   1.90    &  1.90   &  1.2  &  802    \\ 
Ga  	&   1.23  &  1.70    &   1.90    &  1.90   &  1.3  &  801    \\ 
Ge	&   1.18  &  1.70    &   1.90    &  1.90   &  1.3  &  807    \\
As  	&   1.14  &  1.54    &   2.20    &  1.90   &  1.3  &  613    \\ 
Se  	&   1.10  &  1.40    &   2.30    &  1.90   &  1.3  &  571    \\ 
Cd 	&   1.70  &  1.90    &   2.10    &  1.90   &  1.3  &  657    \\ 
In    	&   1.66  &  2.10    &   2.30    &  1.90   &  1.3  &  582    \\ 
Sb  	&   1.56  &  1.90    &   2.30    &  1.90   &  1.2  &  561    \\ 
Te   	&   1.51  &  1.82    &   2.40    &  1.90   &  1.4  &  617    \\ 
\end{tabular}
\end{ruledtabular}
\end{table}

For the $GW$ calculations presented here, all PAW potentials  apply
``approximately'' norm-conserving partial waves (a small norm violation
of up to 10--20\% was allowed for some cases). 
As discussed in the previous section, such potentials should yield an accurate description of
the correlation energy and self-energy even at very long wave vectors,
corresponding to rapid variations in space.
To construct these ``approximately'' norm-conserving partial waves the
following strategies were adopted.
(i) First, non-norm-conserving pseudo partial waves were used,
and it was attempted to set the core radius such that
the pseudo partial wave $\tilde \phi_\alpha(r)$ possesses the same norm as the
all-electron partial wave $\phi_\alpha(r)$ within a sphere around the atom (see Eq.~\ref{equ:PAW}). 
(ii) If this yielded too hard potentials or too large core radii, a norm-conserving
pseudo partial wave $\tilde \phi_\alpha(r)$ was chosen. 
Table~\ref{tab:S2} reports the final core radii for all considered elements. 
Generally small core radii (below 1.6 a.u.) indicate that
option (i) was chosen, whereas larger core radii indicate
that option (ii) was used. 
The same potentials have been
also applied in our recent work using vertex corrected $GW$ calculations.\cite{grueneis2014}

The most shallow core states were treated as valence for all elements, except for boron, carbon, 
nitrogen, oxygen, fluorine, and sulfur. 
In these cases, the core states are well below 10~Ry, and they are very well localized.
Unfreezing them yielded very small changes in test calculations.
For $3d$ elements, as well as Ga, Ge, As, and Se, the  $3s$, $3p$, and $3d$ states 
were treated as valence 
and for $4d$ elements, as well as In, Sb, and Te, the
$4s$, $4p$, and $4d$ states were treated as valence. 
Usually, 3 partial waves were used for each angular quantum number $l$. 
One partial wave was placed at the uppermost core state, one at the binding energy of the valence state 
in the atom,  and a third projector about 20 Ry above the vacuum level. 
This guarantees excellent high energy scattering properties up to about 30 Ry. 
Beyond 30 Ry, however, no projectors and partial waves are available 
so that plane waves beyond 30 Ry are not properly picked up by the projectors $p_\alpha$.
These are the energies where norm-conservation should help.

The plane wave cut-off was chosen to be the maximum $E_{\rm cut}^{pw}$ of all
elements in the considered material. 
To determine basis set converged $GW$ values, the plane wave cut-off $E_{\rm cut}^{pw}$ was systematically 
increased by a factor of 1.25 and 1.587, 
corresponding to an increase of the number of plane waves $N_{pw}$ by a factor 1.5 and 2, respectively.
The results were then extrapolated to the infinite basis set limit, assuming that
the QP energies converge like $(1/E_{\rm cut}^{pw})^{3/2}= 1/N_{pw}$.
For sufficiently large plane wave cut-offs, this behavior was strictly observed for
all materials.
However,  as shown below for the example of ZnO, 
this behavior sets in only at very high plane wave cut-offs for $3d$ states. 
This implies that the $3d$ states might have an error of about 100 meV (errors are smaller 
for $4d$ and also  for the somewhat problematic $2p$ states).
The cut-off for the response function was set to half the plane wave cut-off for the orbitals. 
Correct asymptotic convergence was only observed if both, the plane wave cut-offs for the 
orbitals and the response function, were increased simultaneously.

To  reduce the computational cost, corrections
due to the basis set errors were determined using $3\times 3 \times 3$
$k$-points, and added to the values calculated using $4\times 4 \times 4$
and $6\times 6 \times 6$ $k$-points.

The $GW$ calculations presented here have been performed using the
fully frequency dependent version as described in Refs.~\onlinecite{shishkin2006} and~\onlinecite{shishkin2007}.
For the $G_0W_0$ results presented here, the QP energies were determined
by linearization of the self-energy around the original LDA eigenvalue $E_{n\mathbf{k}}^{0}$
\begin{equation}
\label{equ:qpe}
\begin{split}
 E_{n\mathbf{k}}^{QP} & =  E_{n\mathbf{k}}^{0}+Z_{n\mathbf{k}}\times   \\
&  {\rm Re}[\langle \psi_{n\mathbf{k}}|T+V_{n-e}+V_{H} + \Sigma(E_{n\mathbf{k}}^{0}) | \psi_{n\mathbf{k}}\rangle-  
E_{n\mathbf{k}}^{0}],
\end{split} 
\end{equation}
where  the renormalization factor $Z_{n\mathbf{k}}$ is calculated as
\begin{equation}
\label{equ:z}
   Z_{n\mathbf{k}}= (1-
  { {\rm Re} \langle \psi_{n\mathbf{k}}|\frac{\partial}{\partial\omega}\Sigma(\omega)\Bigl|_{E_{n\mathbf{k}}^{0}} | \psi_{n\mathbf{k}}\rangle)}^{-1}.
\end{equation}
For the  $GW_0$ case, the independent particle screening $\chi_0$ that is used to determine $W_0= v + v \chi_0 v + ...$
is kept fixed at the level of DFT, but the eigenvalues in the
Green's function are updated until convergence is reached. 
In practice, good self-consistency is reached after 4 iterations. 
Note, that we also update the occupancies and Hartree potential in the course of the iterations,
which changes the results slightly for those materials where the band gap is
originally inverted in the DFT calculations (see below).

The final technical detail concerns the treatment of the augmentation 
term on the r.h.s. of Eq.~(\ref{equ:aug1}). It  is approximated by pseudized augmentation functions leading to\cite{kresse99}
\begin{equation}
\begin{split}
\label{equ:aug2}
\langle a | r | i \rangle & = \psi_a^*(r) \psi_i(r) \approx \\ \tilde \psi_a^*(r) \tilde \psi_i(r)
 &+ \sum_{\alpha\beta} \langle \tilde a | p_\alpha \rangle \tilde Q_{\alpha\beta}(r)  \langle p_\beta | \tilde j \rangle\,.
\end{split}
\end{equation}
The  pseudized augmentation density $\tilde Q_{\alpha\beta}(r)$ is evaluated in real space 
and constructed to closely reproduce the exact all electron quantity. 
This is achieved by imposing two constrains:
(i) $\tilde Q_{\alpha\beta}(r)$ must reproduce the multipoles of the exact augmentation charge density $Q_{\alpha\beta}(r)= \phi^*_\alpha(r)\phi_\beta(r)  - \tilde \phi^*_\alpha(r)\tilde \phi_\beta(r)$, 
and (ii) the Fourier transform of $Q_{\alpha\beta}$ and $\tilde Q_{\alpha\beta}$  have to match 
up to a chosen plane wave component. 
The second condition (ii) is not usually used in groundstate calculations,  but employed in 
the Vienna ab initio simulation package (VASP) in QP or any other correlated calculation.

\section{Results}

\subsection{The example of ZnO, GaAs, and AlAs}
\label{sec:ZnO}
\begin{figure}
    \begin{center}
    \includegraphics[width=8cm,clip=true]{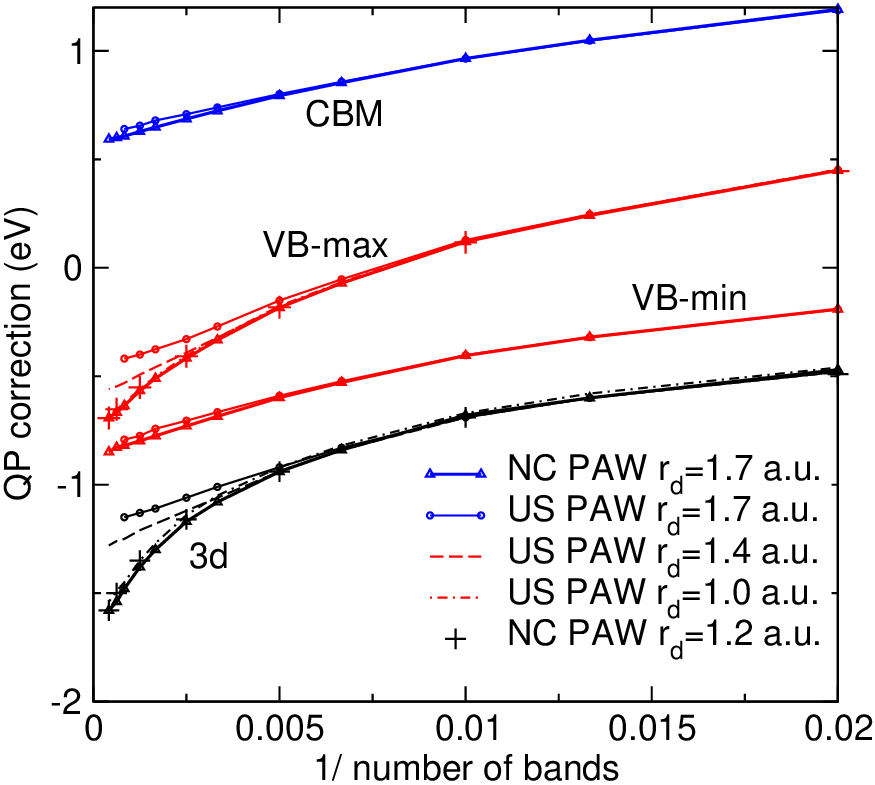}
    \end{center}
    \caption{
   (color online) QP corrections to Kohn-Sham eigenvalues for ZnO $G_0W_0$
 calculations.  The number of bands varies between 50 (right most point)
 and 2400 (left most point). 
 To improve the presentation, an upward shift of  1~eV  was added to the VB  and $3d$ QP corrections
({\rm i.e.}  the true corrections can be obtained by subtracting 1~eV, moving the
 lines down by 1~eV).}
    \label{fig:ZnO}
\end{figure}

\begin{figure}
    \begin{center}
    \includegraphics[width=8cm,clip=true]{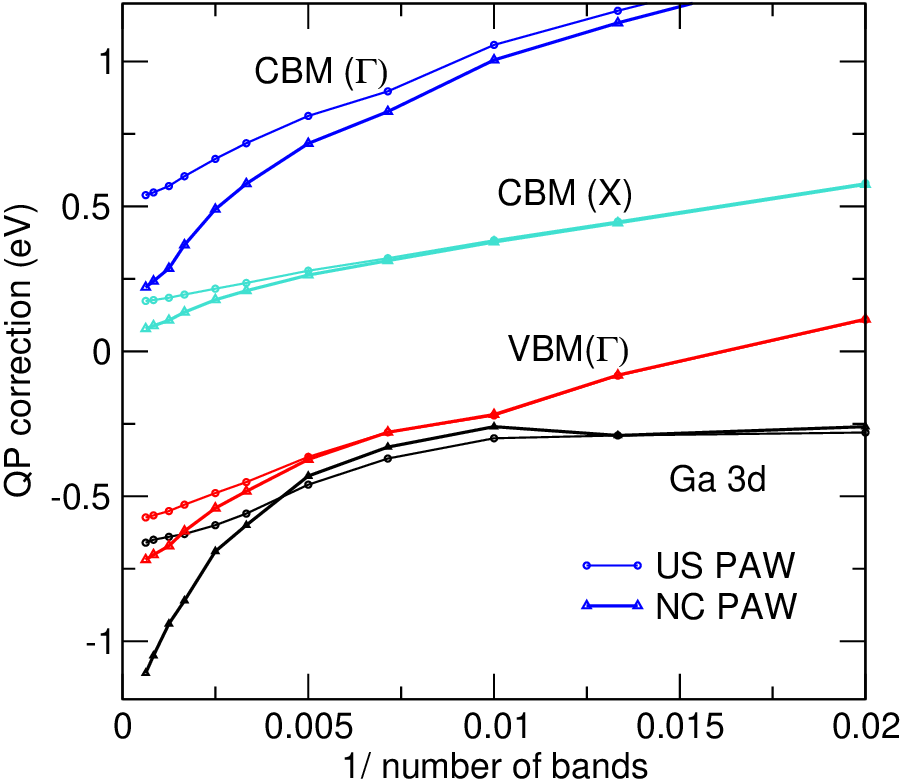}
    \end{center}
    \caption{
   (color online) QP corrections to Kohn-Sham eigenvalues for GaAs $G_0W_0$
 calculations.  The number of bands varies between 50 (right most point)
 and 1600 (left most point). 
 To the $3d$ QP corrections, 1.5~eV has been added.}
    \label{fig:GaAs}
\end{figure}

\begin{figure}
    \begin{center}
    \includegraphics[width=8cm,clip=true]{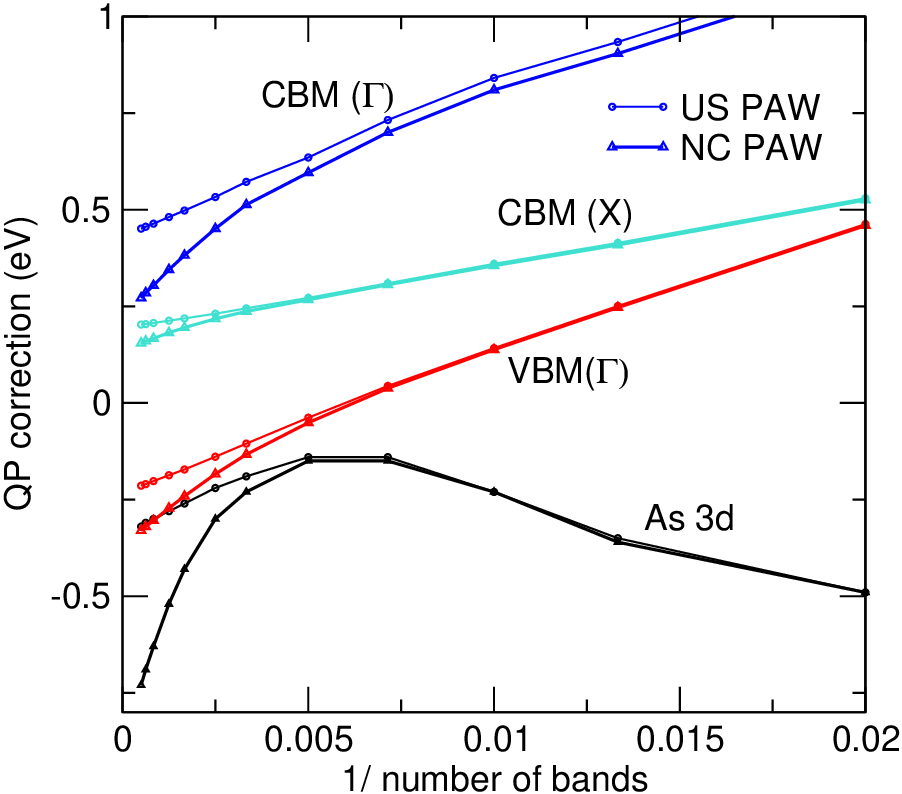}
    \end{center}
    \caption{
   (color online) QP corrections to Kohn-Sham eigenvalues for AlAs $G_0W_0$
 calculations.  The number of bands varies between 50 (right most point)
 and 1600 (left most point). 
 To the VB and $3d$ QP corrections, 0.5~eV and 3~eV has been added, respectively.}
    \label{fig:AlAs}
\end{figure}

As an illustrative example, we  consider ZnO in the zinc-blende
structure. In this section, the k-point set is set to $4\times 4 \times 4$
points to simplify the computationally rather expensive convergence
tests. Furthermore, the plane wave cut-off for the response function
was set to a kinetic energy of 1000 eV, corresponding to a basis set
of about 1750 plane waves for the response function.
We now focus on the convergence of the results with respect to the 
number of orbitals included in the calculation of the Green's function $G$ and
the response function $\chi_0$. 
Fig.~\ref{fig:ZnO} shows the QP corrections of the Kohn-Sham eigenvalues versus 
the inverse of the total number of bands. 

We  first concentrate on the results for the norm-conserving (NC) 
and ultra-soft (US) PAW potentials with a core radius of 1.7~a.u. 
It is clear, that the results are indistinguishable for up to 150--200 bands, 
but beyond that the QP correction of the $3d$~orbitals deviates significantly for
both calculations, with a sharp drop being visible
for the NC PAW potential around 200 orbitals. 
The final converged $d$~band correction differs
by almost 1~eV between the potentials, although both potentials
are practically indistinguishable for groundstate calculations
and $GW$ calculations with few hundreds of orbitals.
The error is related to the lack of projectors at 
high energies, as confirmed by inspection of $\langle p_\alpha | a \rangle$ 
for high energy states $a$, compare Eq.~(\ref{equ:PAW}). 
A standard way to reduce the error of US PAWs is to decrease the core radius used 
upon creation of the PAW potential.
Reducing the core radius for the $d$~electrons $r_d$ to 1.4~a.u. improves
the results, but only slightly, and reasonable agreement with the NC PAW results is
only obtained at a much smaller core radius of $r_d$=1.0~a.u.,
where the violation of the norm is only about 20~\%.
At this point, the groundstate calculations with the US PAW $r_d$=1.0~a.u. potential, however, require
already a larger basis set than the calculations with the NC PAW $r_d$=1.7~a.u. potential. 
Specifically, the required plane wave energy cut-offs for accurate groundstate
calculations are 1150 eV for the NC PAW $r_d$=1.7~a.u.,
and 500, 700, and 1400 eV for the US PAW potentials with core radii of
$r_d$=1.7, 1.4, and 1.0~a.u. respectively.
Furthermore, within the line thickness, 
the results for NC PAW potentials are 
independent of the core radius for radii between 
1.9~a.u. and 1.2~a.u. (see crosses in Fig.~\ref{fig:ZnO}). 
Hence, the remaining Zn calculations were performed for
a softer NC potential with the core radius increased to 1.9~a.u.
(compare Tab.~\ref{tab:S2}).

The errors in the QP corrections are most pronounced for the $3d$ states. 
However, the valence band  maximum (VBM), which is
dominated by oxygen $2p$ states follows the $3d$ states
partially. 
This is of course related to the strong covalent
Zn-$3d$ O-$2p$ interaction: at the $\Gamma$ point
the Zn-$3d$ states hybridize with the O-$2p$ states 
forming Zn-$3d$ O-$2p$ bonding  and antibonding 
linear combinations. 
Since the antibonding linear combination, which forms the threefold 
degenerated valence band maximum at $\Gamma$, has a strong Zn-$3d$ character, 
the states partly follow the behavior of the $3d$ states. 
As a consequence, when the number of orbitals increases the band gap opens.

This is however not always the case, as exemplified for
GaAs and AlAs  (Fig. \ref{fig:GaAs} and Fig. \ref{fig:AlAs}).
For these materials, there is no (AlAs) or little hybridization
between the $3d$ states and the valence and conduction band states.  
Nevertheless, the $3d$ states still influence the
valence band and conduction band states, as can be understood from Eq.~(\ref{equ:exactGW}).
The $d$ electrons will result in a spherically symmetric density at
the As and Ga atom, and this causes an attractive Coulomb hole at the atomic sites. 
The localized $d$ states themselves and the $s$ like states are most 
strongly affected by this local potential and pulled down to lower energies.

For GaAs,  the first thing to note is that the Ga $3d$ states first shift slightly upwards 
in energy and then suddenly drop in energy. 
This is a result of the strong spatial localization of the $3d$ electrons in Ga. 
The effect is even more pronounced for the As $3d$ states as shown in Fig.~\ref{fig:AlAs}. 
The more important observation, however, is that the valence band states at the $\Gamma$ point
and the $s$ like conduction band minimum CBM($\Gamma$) follow this drop. 
The CBM at the X point exhibits almost no slope, much smaller than that
of the valence band maximum VBM($\Gamma$).
This behavior reduces the $\Gamma-\Gamma$ transition, but increases
the $\Gamma-$X transition. 
Obviously increasing the basis set can have quite a substantial impact on the relative position of the 
conduction band states and thus the direct and indirect gaps. 

Remarkably, similar effects are observed even for AlAs as shown in Fig.~\ref{fig:AlAs}.
The QP corrections are about a factor of two smaller, since the coupling
to the strongly localized As $3d$ states is now rather small
(note the different $y$ scale in  Fig.~\ref{fig:AlAs}). 
Again the QP correction of the CBM($\Gamma$) state exhibits the largest slope, 
the CBM(X) the smallest one, and the VBM($\Gamma$) is in between.
Consequently the indirect gap of AlAs increases, when the basis set
size is increased.
Clearly, whenever accurate direct and indirect gaps  are needed
the $d$ electrons must be taken into account, and this applies
to both the anion  as well as the cation  (As, Se, Te).
Furthermore,  PAW potentials without norm-conserving partial
waves yield too small QP corrections at $\Gamma$, in particular, for the CBM($\Gamma$) state, as indicated
by the thin lines in Fig.~\ref{fig:GaAs} and~\ref{fig:AlAs}, and consequently too large
direct gaps and too small indirect gaps.

We performed extensive tests for ZnO, AlAs, CdS, GaN, and InP, 
all confirming these observations. 
Although up to about 100--200 orbitals per atom, the results are independent of the choice of
the potential, once the number of orbitals is increased to several 1000, significant deviations
between the QP corrections become discernible. 
US PAW potentials always yield too small QP corrections for the $d$~states and the results converge
only slowly with decreasing core radii to a limiting value.
In contrast, the results are almost independent of the specific choice of
the core radius for NC PAW potentials.

A final word of caution is in place here. In the present
implementation, the code restores an approximation of the exact all electron
density on the plane wave grid. 
As briefly hinted at after Eq.~(\ref{equ:aug2}), this is achieved by performing 
a Fourier transformation of the exact augmentation charge density
$Q_{\alpha\beta}(r)= \phi^*_\alpha(r)\phi_\beta(r)  - \tilde \phi^*_\alpha(r)\tilde \phi_\beta(r)$
to reciprocal space $Q_{\alpha\beta}(G)$ and then expanding
the augmentation density in a set of orthogonal functions
localized at each atomic site. 
In the present calculations, three functions for each quantum number $l$ are used to expand the
density difference between the all electron and pseudo partial waves
({\tt NMAXFOCKAE = 2} in VASP). 
They are used to restore the proper norm ($l=0$), dipoles ($l=1$), and quadrupoles ($l=2$)
and higher multipoles ($l>2$).
The first function serves to restore the exact $Q_{\alpha\beta}(G)$ around $G=0$
and is not required if the potentials are norm-conserving. 
The other two functions are chosen to restore the density
at larger plane wave vectors $G$.
Since the method guarantees that
the density on the plane wave grid is almost exactly equivalent
to the exact all electron density up to plane waves corresponding
to a kinetic energy of 400 eV,  the fairly complicated one-center terms in the screening 
matrix $\chi_0$ and $W$ do not need to be implemented.
We believe that the present procedure and implementation is very accurate, since
it yields identical QP corrections for all potentials, if the number of unoccupied orbitals is not too large
(right hand side in Figs.~\ref{fig:ZnO}, \ref{fig:GaAs}, and~\ref{fig:AlAs}).
With norm-conserving partial waves the results become even  robust and stable for
very large basis sets.
Restoring just the norm of $Q_{\alpha\beta}(r)$ was
found to be insufficient for $GW$ calculations.
If only the norm is restored in our code, differences in the QP correction for 
different potentials can be around 1~eV for localized $d$~states, even when only
a small number of unoccupied orbitals is included in the $GW$ calculations.

Finally, the interaction between the core and valence
electrons is always evaluated exactly at the level of HF without
any shape approximation, which is not always
the case for codes using standard norm-conserving potentials. 
Based on our experience, we would not expect that such standard potentials without 
the PAW information (all-electron orbitals) can yield similar accuracy.

Technically, we believe that the procedures adopted here allow
to get accurate and converged results for
virtually any material, although the construction of PAW
potentials can be tedious at times. For instance ghost-states
at energies below bound states or in the conduction band
need to be avoided and sufficient number of projectors
at high energies need to be included.
 
\begin{figure}
    \begin{center}
    \includegraphics[width=8.5cm,clip=true]{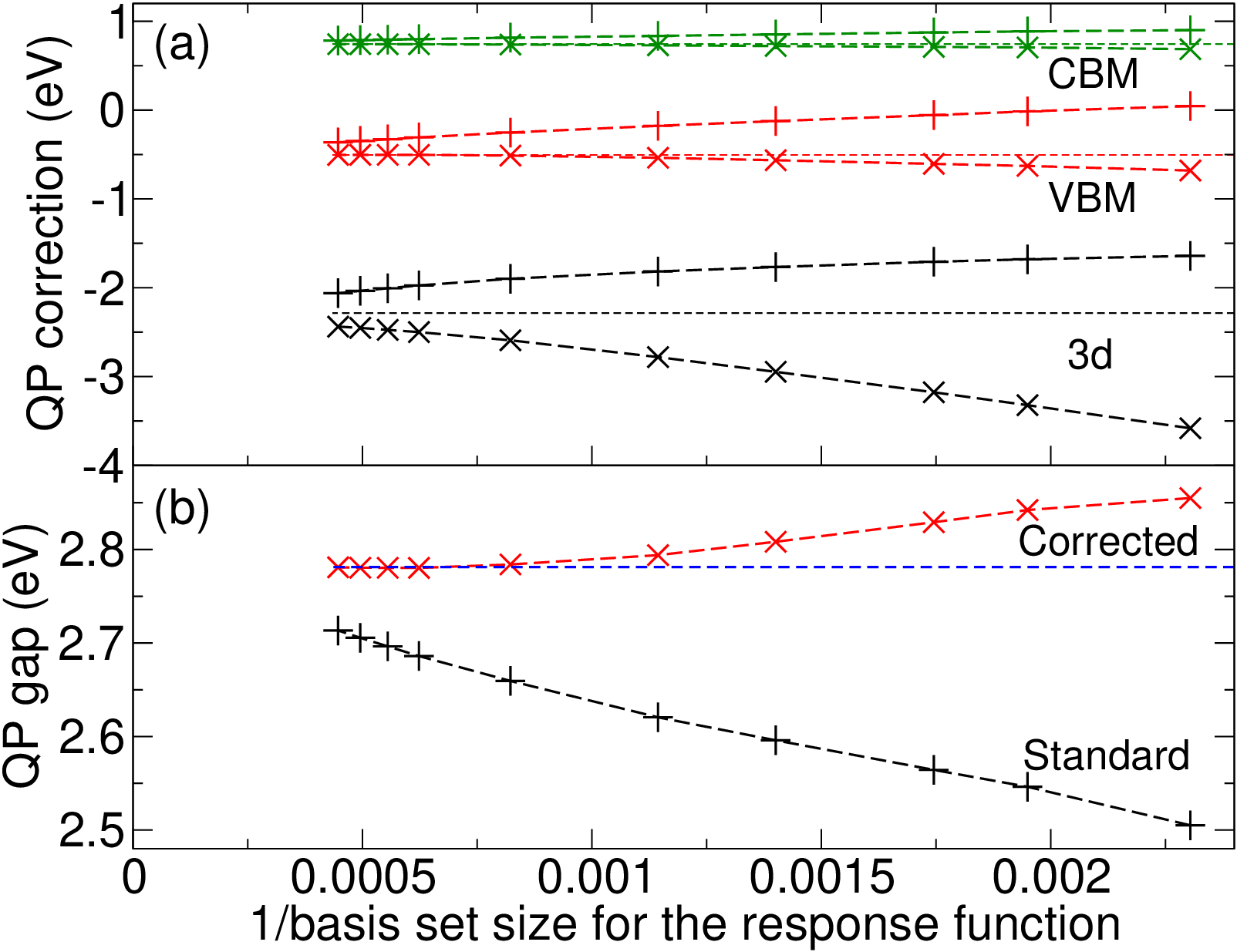}
    \end{center}
    \caption{
 Dependence of (a) the unscaled QP correction (without $Z$) and (b) of the quasi-particle
band gap of wurzite ZnO on the basis set used for the response function.
Bare data ($+$ sign) and data corrected for the leading error according to Eq.~(\ref{equ:exactGWN}),
marked with $\times$ sign, are shown.
Horizontal lines are extrapolated values. In (a)
the data for the $3d$ and VBM states were shifted to a higher energy by 1.5~eV.}
    \label{fig:ZnOacc}
\end{figure}

\subsection{Basis set incompleteness correction}

Converging $GW$ results with respect to the basis set size 
can become computationally prohibitively expensive and
can require tedious extrapolation procedures, as performed here or in other work.\cite{friedrich2011,klimes2014OEP}
An alternative to the extrapolation is to use the simple basis set incompleteness correction
for the QP energies, Eq.~(\ref{equ:exactGWN}), and here we test how accurate this is.
For $GW$ calculations, it is most efficient to correct for the incompleteness of the
response function basis set (limited number of $N_{\rm pw}^\chi$).
In this case, it is important to converge the response function with respect to the number of bands,
by using a sufficiently large cut-off $E_{\rm cut}^{\rm pw}$ for the orbitals and using all available bands.

To test the correction, we again consider the well investigated ZnO  here in the wurzite  structure
(see, e.g., Refs.~\onlinecite{shih2010} and \onlinecite{friedrich2011}).
Figure~\ref{fig:ZnOacc} (a) shows the dependence of the QP corrections on the number of basis
functions in the response function both without and with the correction according to Eq.~(\ref{equ:exactGWN}).
The thin dashed lines show values extrapolated from the last three (non-corrected) data points assuming a linear
convergence with the inverse of the basis set size, i.e. $\varepsilon(N_{\rm pw}^\chi)=\varepsilon(\infty)+A/N_{\rm pw}^\chi$.
The renormalization factor $Z$ was not applied, and the plane wave cut-off
for orbitals was taken to be three times the cut-off for the response properties, which was between 250 and 750~eV.
Clearly the QP corrections of the VB maximum and the CB minimum converge much faster when the correction is applied.
The correction is larger for the VB maximum, which contains a contribution of the 3$d$ states of Zn.
The corrected shifts of the 3$d$ states deviate initially more from the extrapolated value, and the difference is still pronounced
even for the largest basis set used.
However, the extrapolated value  for the 3$d$ states might still posses some error because of the very slow convergence
with $N_{\rm pw}^\chi$.
Importantly, the coefficient for the correction calculated using Eq.~(\ref{equ:exactGWN}) can be directly compared to the
value of the coefficient $A$ obtained from the fit.
For the VB maximum and CB minimum the agreement is excellent, the fit yields $A=317$~eV and 89~eV for the VB maximum and CB minimum, respectively,
while Eq.~(\ref{equ:exactGWN}) gives $A=315$~eV and 93~eV.
For the 3$d$ state, we obtain 502~eV from the fit and 842~eV from Eq.~(\ref{equ:exactGWN}). 
Clearly even larger cut-offs would have to be used to observe the $1/N_{\rm pw}^\chi$ behavior.
However, this is hardly possible, not least because the largest calculation we performed used over 12000 unoccupied bands.

The faster convergence of the QP corrections leads to a much faster convergence of the band gap, shown in Figure~\ref{fig:ZnOacc} (b).
When the correction is used, a rather small cut-off of $E_{\rm cut}^{\chi}=300$~eV (corresponding to about 600 plane waves in the response function)
is sufficient to obtain a band gap converged to within 50~meV.
For the same settings the uncorrected band gap is underestimated by more than 200~meV.
Even in the largest calculation we performed, using $E_{\rm cut}^{\chi}=750$~eV, the gap is still underestimated by $\approx$70~meV.
To obtain similar error from the corrected calculation, it is sufficient to set the cut-off to a much smaller value,
$E_{\rm cut}^{\chi}=250$~eV. 
With similar errors, the corrected calculations are about 30 times cheaper.
Clearly, the correction is a promising tool to improve the convergence of the QP energies
in $GW$ calculations. 
We also note that this correction is complementary to convergence accelerations 
with respect to the number of included orbitals.\cite{bruneval2008,berger2012}
Here, we correct for the error
caused by the finite basis set used for the response function, whereas those address
errors incurred by truncating the virtual orbital set.

\subsection{$G_0W_0$ band gaps for semiconductors and insulators}

\begin{table}[t]
\caption{
Position of valence band (VB) maximum at X, L (with respect to $\Gamma$),
and conduction band (CB) minimum at $\Gamma$, X, and L with respect
to VB maximum at $\Gamma$ for the local density approximation. 
For materials with an inverted gap, the VB maximum is
set to the triple degenerated state that 
normally corresponds to the VB maximum.
All materials were considered in the diamond or zinc blende structure, except
for GaN and ZnO where the lines ``wz" report the results for the
wurzite structures.
}
\label{tab:LDA}
\begin{ruledtabular}
\begin{tabular}{lrrrrrrrr}
            & $\Gamma_{\rm VBmin}$  &  $\Gamma_c$ & L$_v$ & L$_c$ & X$_v$ & X$_c$ &  $d$  \\ 
\hline   
C           & -21.32 &    5.54 &   -2.79 &    8.38 &   -6.29 &    4.70 &  \\
SiC         & -15.33 &     6.28 &   -1.06 &    5.38 &   -3.20 &    1.30 &  \\
Si          & -11.96 &    2.52 &   -1.20 &    1.42 &   -2.85 &    0.60 &  \\
Ge          & -12.81 &   -0.15 &   -1.41 &    0.06 &   -3.08 &    0.66 &  \\
BN          & -20.07 &    8.68 &   -1.94 &   10.19 &   -4.91 &    4.34 &  \\
AlP         & -11.50 &    3.10 &   -0.77 &    2.66 &   -2.12 &    1.44 &  \\
AlAs        & -11.90 &    1.86 &   -0.82 &    2.02 &   -2.17 &    1.35 &  \\
AlSb        & -10.77 &    1.46 &   -0.91 &    1.26 &   -2.19 &    1.15 &  \\
GaN         & -15.64 &    1.63 &   -0.95 &    4.42 &   -2.66 &    3.24 &  -13.50 \\
wz          &        &    1.94 &         &         &         &         &  -13.30  \\
GaP         & -12.51 &    1.63 &   -1.13 &    1.51 &   -2.71 &    1.49 &  -14.68 \\
GaAs        & -12.76 &    0.32 &   -1.15 &    0.86 &   -2.69 &    1.34 &  -14.91 \\
GaSb        & -11.59 &   -0.06 &   -1.17 &    0.32 &   -2.61 &    0.83 &  -15.14 \\
InP         & -11.52 &    0.48 &   -0.98 &    1.31 &   -2.35 &    1.60 &  -14.13 \\
InAs        & -11.88 &   -0.43 &   -0.99 &    0.79 &   -2.33 &    1.43 &  -14.30 \\
InSb        & -10.77 &   -0.38 &   -1.01 &    0.42 &   -2.27 &    1.25 &  -14.51 \\
MgO         & -17.00 &    4.68 &   -0.67 &    7.75 &   -1.37 &    8.89 &         \\
ZnO         & -17.68 &    0.62 &   -0.80 &    5.32 &   -2.21 &    5.13 &   -5.30 \\
wz          &        &    0.75 &         &         &         &         &   -5.24 \\
ZnS         & -13.08 &    1.87 &   -0.87 &    3.10 &   -2.23 &    3.19 &   -6.31 \\
ZnSe        & -13.28 &    1.05 &   -0.87 &    2.36 &   -2.20 &    2.81 &   -6.55 \\
ZnTe        & -11.82 &    1.06 &   -0.90 &    1.65 &   -2.18 &    2.17 &   -6.94 \\
CdO         & -15.73 &    0.92 &    1.42 &    5.68 &   -0.98 &    5.10 &   -5.14 \\
CdS         & -12.40 &    0.90 &   -0.78 &    2.79 &   -1.95 &    3.30 &   -7.60 \\
CdSe        & -12.65 &    0.36 &   -0.77 &    2.19 &   -1.89 &    2.94 &   -7.80 \\
CdTe        & -11.23 &    0.55 &   -0.79 &    1.66 &   -1.89 &    2.45 &   -8.18 
\end{tabular}
\end{ruledtabular}
\end{table}

\begin{table}[t]
\caption{
$G_0W_0$@LDA results for absolute shift of the valence band (VB) at $\Gamma$ compared to LDA calculations ($\Delta$ IP),
position of the VB minimum ($\Gamma_{\rm VBmin}$),
position of VB maximum at X, L (with respect to $\Gamma$),
and conduction band (CB) minimum at $\Gamma$, X, and L. For Ga, In, Zn, and Cd, the position of
the $d$ band (average at the $\Gamma$ point) is also indicated.
Calculations are for $6\times 6 \times 6$ k-points. 
}
\label{tab:G0W0}
\begin{ruledtabular}
\begin{tabular}{lrrrrrrrr}
             
            &$\Delta$IP     & $\Gamma_{\rm VBmin}$  &  $\Gamma_c$ & L$_v$ & L$_c$ & X$_v$ & X$_c$ &  $d$  \\ 
\hline   
C           &   -1.07 &  -22.06 &    7.43 &   -2.92 &   10.37 &   -6.55 &    6.21 &         \\
SiC         &   -0.92 &  -15.61 &    7.30 &   -1.09 &    6.58 &   -3.27 &    2.43 &         \\
Si          &   -0.60 &  -11.82 &    3.21 &   -1.21 &    2.06 &   -2.86 &    1.22 &         \\
Ge          &   -0.54 &  -12.69 &    0.43 &   -1.42 &    0.56 &   -3.09 &    1.10 &         \\
BN          &   -1.35 &  -21.09 &   11.29 &   -2.08 &   12.25 &   -5.19 &    6.37 &         \\
AlP         &   -0.83 &  -11.36 &    4.11 &   -0.78 &    3.70 &   -2.13 &    2.42 &         \\
AlAs        &   -0.82 &  -11.81 &    2.82 &   -0.83 &    2.96 &   -2.17 &    2.23 &         \\
AlSb        &   -0.65 &  -10.65 &    2.27 &   -0.91 &    1.99 &   -2.20 &    1.82 &         \\
GaN         &   -1.04 &  -15.86 &    2.88 &   -0.97 &    5.95 &   -2.68 &    4.59 &  -15.87 \\
wz          &   -1.08 &         &    3.23 &         &         &         &         &  -15.66 \\
GaP         &   -0.70 &  -12.34 &    2.50 &   -1.14 &    2.36 &   -2.71 &    2.25 &  -16.86 \\
GaAs        &   -0.63 &  -12.59 &    1.08 &   -1.15 &    1.57 &   -2.68 &    1.96 &  -17.10 \\
GaSb        &   -0.54 &  -11.45 &    0.54 &   -1.17 &    0.84 &   -2.61 &    1.29 &  -17.32 \\
InP         &   -0.61 &  -11.32 &    1.13 &   -0.99 &    2.00 &   -2.35 &    2.21 &  -15.73 \\
InAs        &   -0.60 &  -11.69 &    0.13 &   -1.00 &    1.43 &   -2.33 &    2.02 &  -15.84 \\
InSb        &   -0.54 &  -10.61 &    0.13 &   -1.03 &    0.93 &   -2.28 &    1.72 &  -16.06 \\
MgO         &   -2.01 &  -17.47 &    7.55 &   -0.71 &   10.86 &   -1.44 &   11.91 &         \\
ZnO         &   -1.53 &  -17.75 &    2.46 &   -0.79 &    7.43 &   -2.15 &    7.00 &   -6.22 \\
wz          &   -1.78 &         &    2.83 &         &         &         &         &   -6.09 \\
ZnS         &   -1.15 &  -12.62 &    3.36 &   -0.86 &    4.70 &   -2.18 &    4.62 &   -7.55 \\
ZnSe        &   -1.09 &  -12.97 &    2.38 &   -0.85 &    3.75 &   -2.12 &    4.06 &   -7.82 \\
ZnTe        &   -0.84 &  -11.59 &    2.17 &   -0.88 &    2.71 &   -2.13 &    3.15 &   -8.31 \\
CdO         &   -0.81 &  -16.16 &    1.57 &    1.14 &    6.83 &   -0.98 &    6.46 &   -6.78 \\
CdS         &   -0.96 &  -11.98 &    2.05 &   -0.77 &    4.09 &   -1.91 &    4.48 &   -8.67 \\
CdSe        &   -0.90 &  -12.37 &    1.33 &   -0.75 &    3.27 &   -1.83 &    3.93 &   -8.84 \\
CdTe        &   -0.75 &  -10.99 &    1.46 &   -0.78 &    2.58 &   -1.85 &    3.28 &   -9.34
\end{tabular}
\end{ruledtabular}
\end{table}

\begin{table}[t]
\caption{
Same as Table \ref{tab:G0W0} but for $G_0W_0$@PBE. The IPs are
generally 0.15~eV more negative than for $G_0W_0$@LDA.
}
\label{tab:G0W0_PBE}
\begin{ruledtabular}
\begin{tabular}{lrrrrrrrr}
             
            &$\Delta$IP     & $\Gamma_{\rm VBmin}$  &  $\Gamma_c$ & L$_v$ & L$_c$ & X$_v$ & X$_c$ &  $d$  \\ 
\hline   
C           &   -1.22 &  -22.12 &    7.43 &   -2.93 &   10.38 &   -6.58 &    6.23 &   \\
SiC         &   -1.05 &  -15.69 &    7.35 &   -1.10 &    6.62 &   -3.30 &    2.42 &   \\
Si          &   -0.72 &  -11.83 &    3.25 &   -1.21 &    2.14 &   -2.86 &    1.28 &   \\
Ge          &   -0.82 &  -12.70 &    0.63 &   -1.43 &    0.67 &   -3.08 &    1.16 &   \\
BN          &   -1.53 &  -21.05 &   11.33 &   -2.06 &   12.29 &   -5.18 &    6.40 &   \\
AlP         &   -0.96 &  -11.37 &    4.23 &   -0.77 &    3.79 &   -2.13 &    2.48 &   \\
AlAs        &   -1.01 &  -11.82 &    2.99 &   -0.82 &    3.08 &   -2.17 &    2.31 &   \\
AlSb        &   -0.84 &  -10.67 &    2.40 &   -0.91 &    2.07 &   -2.20 &    1.87 &   \\
GaN         &   -1.20 &  -15.93 &    2.85 &   -0.98 &    5.93 &   -2.70 &    4.50 &  -15.82 \\
wz          &   -1.23 &         &    3.20 &         &         &         &         &  -15.61 \\
GaP         &   -0.86 &  -12.34 &    2.62 &   -1.14 &    2.45 &   -2.71 &    2.30 &  -16.77 \\
GaAs        &   -0.91 &  -12.56 &    1.23 &   -1.14 &    1.68 &   -2.66 &    2.04 &  -16.97 \\  
GaSb        &   -0.77 &  -11.41 &    0.68 &   -1.17 &    0.92 &   -2.60 &    1.34 &  -17.22 \\
InP         &   -0.78 &  -11.29 &    1.23 &   -0.99 &    2.10 &   -2.35 &    2.28 &  -15.66 \\
InAs        &   -0.83 &  -11.65 &    0.23 &   -1.00 &    1.48 &   -2.32 &    2.04 &  -15.73 \\ 
InSb        &   -0.75 &  -10.57 &    0.25 &   -1.03 &    0.99 &   -2.28 &    1.76 &  -15.94 \\
MgO         &   -2.16 &  -17.61 &    7.55 &   -0.73 &   10.80 &   -1.45 &   11.82 &  \\
ZnO         &   -1.73 &  -17.84 &    2.42 &   -0.79 &    7.41 &   -2.17 &    6.93 &   -6.16 \\  
wz          &   -1.87 &         &    2.76 &         &         &         &         &   -5.98 \\ 
ZnS         &   -1.33 &  -12.64 &    3.46 &   -0.87 &    4.79 &   -2.19 &    4.68 &   -7.45 \\
ZnSe        &   -1.34 &  -12.95 &    2.55 &   -0.85 &    3.88 &   -2.13 &    4.15 &   -7.64 \\ 
ZnTe        &   -1.07 &  -11.56 &    2.27 &   -0.89 &    2.78 &   -2.14 &    3.21 &   -8.10 \\
CdO         &   -0.95 &  -16.29 &    1.50 &    1.29 &    6.74 &   -0.99 &    6.45 &   -6.20 \\
CdS         &   -1.11 &  -11.97 &    2.15 &   -0.78 &    4.19 &   -1.93 &    4.55 &   -8.62 \\
CdSe        &   -1.15 &  -12.34 &    1.52 &   -0.75 &    3.44 &   -1.84 &    4.06 &   -8.71 \\ 
CdTe        &   -0.99 &  -10.96 &    1.57 &   -0.79 &    2.67 &   -1.86 &    3.35 &   -9.19    


\end{tabular}
\end{ruledtabular}
\end{table}

Tables \ref{tab:LDA}, \ref{tab:G0W0}, and \ref{tab:GW0} collect the band gaps for
the materials considered in the present work for LDA,
$G_0W_0$@LDA, and $GW_0$@LDA calculations starting from LDA orbitals. 
The results for $G_0W_0$ starting from PBE orbitals are shown in Tab.~\ref{tab:G0W0_PBE}.

The first important issue to note is that some of the materials show
band inversion in LDA and PBE, namely  Ge, GaSb, InAs, and
InSb: for these materials the threefold degenerated (cation) $p$  orbitals
are incorrectly above the anion $s$ orbital
at the $\Gamma$ point. 
In Tab.~\ref{tab:LDA}, the gap is then given as a negative value. 
From Table \ref{tab:G0W0}, we see
that the band inversion is already abolished in the first $G_0W_0$ iteration, but
unfortunately the band inversion causes sometimes convergence problems in the 
$GW_0$ iteration, specifically this happened for CdO, where subsequent iterations 
did not converge. 
In these cases, a further technical problem is that both in the $GW_0$, as well as $G_0W_0$
calculations, the screening of the system is practically metallic. 
Hence we expect that these results are possibly not well converged with respect to the k-point mesh, and
the final reported numbers should
be considered with some caution. Another reason for errors is
the frequency integration. We present the polarizability on a discretized frequency
grid,\cite{shishkin2006} with 200 frequency points, and double checked whether a reduction 
of the frequency points from 200 to 100 points changes the results. 
For all systems, except ZnO, the changes in the QP energies and gap are well below 50 meV,
with most changes being only 10--20 meV. 
For ZnO the gap increases by 100 meV, when the number of frequency
points is reduced to 100.
The origin for this change is the one single sharp $s$-$p$ transition
occurring at the $\Gamma$ point, which is not accurately represented
with the coarser frequency grid. 
Error bars for ZnO are, therefore, possibly somewhat larger than for other systems. 

The effect of the starting functional (LDA versus PBE, Tab.~\ref{tab:G0W0} versus~\ref{tab:G0W0_PBE}) 
is small for all materials without $d$ electrons (see, for example Si). 
However, if $d$ electrons are present, 
the $d$ electrons shift slightly towards the Fermi-level by 0.1~eV in the $G_0W_0$,
as well as in the preceding PBE calculations.
More notable is the increase of the band gap
by about 0.2 eV, in particular, for the $\Gamma-\Gamma$ transition
for both PBE (not shown) and $G_0W_0$@PBE.
This implies that $G_0W_0$@LDA calculations can not be straightforwardly compared
to $G_0W_0$@PBE calculations for systems with $d$ electrons. 
Although the PBE starting point seems better
suited, since the final gaps are larger and always in better
agreement with experiment, we will mainly concentrate on the $G_0W_0$@LDA 
calculations in the following discussion, since it is still common practice to
start from LDA orbitals.

Let us start with a discussion of those materials that have been
most widely discussed in literature, Si and ZnO.
For Si, Friedrich~{\it et al.}\cite{friedrich2006}
reported valence band and conduction band positions  
of 3.2 eV ($\Gamma-\Gamma$), $-1.22$, 2.12 ($\Gamma-$L), $-2.92$, 1.19 ($\Gamma-$X)
compared to our values of  3.22 ($\Gamma-\Gamma$),    $-1.21$, 2.06 ($\Gamma-$L) $-2.87$, 1.23 ($\Gamma-$X).
However, Friedrich's calculations were performed using only 250 bands, and they 
also note that ``the results are lowered by 0.02 eV if the k-point
mesh is fully converged, and by another 0.02 eV if screening
due to the $2p$ electrons is included in the correlation self-energy''.
This will improve agreement with our results even further, since we have included
the $2p$ electrons in the valence.

ZnO is certainly the most controversial material, with reported band gaps varying
between 2.1 eV and 3.4 eV for $G_0W_0$ calculations.\cite{shishkin2007,shih2010}
The larger value of 3.4 has been shown to be related to the plasmon pole model,\cite{stankovski2011}
whereas the smaller value of 2.1~eV has been reported for VASP, albeit
for zinc blende and PBE orbitals.\cite{shishkin2007} This decreases the gap by 0.2~eV and 0.1~eV, respectively;
the $G_0W_0$@LDA band gap for wurzite ZnO is 2.4~eV with a similar setup and potential
as in Ref. \onlinecite{shishkin2007} .
In the present calculations, we predict a basis set converged band gap of 2.83~eV
using $6\times 6\times 6$ k-points.
Increasing the grid to $8\times 8\times 8$ k-points, the value increases
to 2.87~eV. These values are only slightly larger then the value of  2.83~eV reported by 
Friedrich~{\it et al.}\cite{friedrich2011} for $8\times 8\times 8$ k-points.
Due to the very slow convergence with the number of orbitals and exceedingly slow k-point convergence, 
we, however, want to emphasize that our ZnO values are possibly less accurate than for
other materials:
errors below $\pm 0.05$~eV are technically very difficult
to achieve for this material. 
The important point, however, is that our present values
are significantly larger than the values we and many other groups
have reported before.\cite{shishkin2007,shih2010,berger2012}
The main reason is certainly insufficient
convergence with respect to the basis sets for the orbitals
or response function, or the use of an incomplete basis set
for the atom-centered basis functions in full potential
methods or projector augmented wave methods.

Concerning the $d$ levels and the other materials considered in the present
work, we will start with  a comparison with the most recent
full potential linearized augmented-plane-wave (FLAPW) values by Li~{\it et al.}\cite{li2012NJP}
For the $d$ states, they report values of $-6.84$, $-7.12$, $-7.55$ (ZnS, ZnSe, ZnTe) and
$-8.16$, $-8.45$, $-8.74$ (CdS, CdSe, CdTe), typically 0.5--0.7~eV higher than our 
present values. 
This is certainly a result of insufficient basis
set convergence, since in the FLAPW calculations the number of bands is nowhere close to
the values required to see the drop in the $d$ level energies
(compare Fig.~\ref{fig:ZnO}). 
Compared to our own previous values
reported in Ref.~\onlinecite{shishkin2007} we note that
these early calculations were neither basis set converged, nor
did we very accurately restore the all electron density distribution on the plane
wave grid. 
Both errors might or might not accidentally cancel; certainly
the present values are more accurate and reliable and
supersede the previous values.

For the band gaps,  we note that the FLAPW  calculations reported
in Ref.~\onlinecite{abal2008} are generally
smaller than our $G_0W_0$ band gaps:
1.00 (Si), 5.42  (C), BN (6.03), AlP (2.18), GaAs (1.29),
compared to the present values of  1.10 (Si), 5.61 (C), BN (6.37),
AlP (2.43), GaAs( 1.07).  We suspect that this
is also a result of insufficient convergence with
respect to the FLAPW basis set inside the atomic  spheres. 
Too small one center basis sets  will yield seemingly
converged results with respect to the plane wave basis set,
although the true band gaps are possibly much larger.
More accurate values  have been published by Friedrich {\it et al.} who reported
carefully converged band gaps for several semiconductors and
insulators also using the FLAPW method.\cite{friedrich2010}
In this case, the one-center basis sets were improved using
local orbitals (LO). They find band gaps of 
1.11 (Si), GaAs (1.29), CdS (2.18), GaN (2.83), 5.62 (C), BN (6.20), MgO (7.17),   most
in good agreement with our values
1.10 (Si), GaAs (1.07), CdS (2.10),  GaN (2.90), 5.61 (C), BN (6.37), MgO (7.54).
As to why our band gaps are generally larger for oxides and nitrides
compared to the FLAPW calculations and our previous calculations in Ref.~\onlinecite{shishkin2007},
we note that the $2p$ levels need to be treated with  similar
care as the $d$ states, and we suspect that the one-center basis
sets for those materials were not fully converged in the FLAPW calculations.
For GaAs, our present band gap is smaller than  the two FLAPW values (both around 1.30).
We believe that the origin of this behavior is insufficient
basis set convergence in both FLAPW calculations, as already discussed above and shown
in Fig.~\ref{fig:GaAs}. 
When the basis set size is increased for GaAs,
the direct gap decreases, since the $s$ like  conduction
band states at $\Gamma$ experience the attractive Coulomb hole of
the $d$ electrons. 
If the basis set is not sufficiently large a too large gap will be predicted.
It, however, also goes without saying that it is possible that one or the other
value in our tables might need revision, when we have
more codes and values to compare to. 
At present, we have done everything to converge our results carefully and
have checked every single PAW potential painstakingly,
typically comparing them to two other PAW potentials with even smaller
core radii.

\begin{table}[t]
\caption{
Same as Table~\ref{tab:G0W0} for $GW_0$@LDA. $GW_0$@LDA calculations did not
converged for CdO because of the incorrect band order for the LDA starting
orbitals. The $GW_0$@PBE are reported in the supplementary material
of Ref.~\onlinecite{grueneis2014}}
\label{tab:GW0}
\begin{ruledtabular}
\begin{tabular}{lrrrrrrrr}
            &$\Delta$IP     & $\Gamma_{\rm VBmin}$  &  $\Gamma_c$ & L$_v$ & L$_c$ & X$_v$ & X$_c$ &  $d$  \\ 
\hline
C           &   -1.21 &  -22.38 &    7.57 &   -3.02 &   10.60 &   -6.71 &    6.36 &  \\
SiC         &   -1.10 &  -15.80 &    7.50 &   -1.10 &    6.79 &   -3.32 &    2.61 &  \\
Si          &   -0.72 &  -11.93 &    3.32 &   -1.22 &    2.15 &   -2.90 &    1.31 &  \\
Ge          &   -0.56 &  -12.78 &    0.53 &   -1.44 &    0.62 &   -3.12 &    1.15 &  \\
BN          &   -1.56 &  -21.30 &   11.66 &   -2.10 &   12.58 &   -5.25 &    6.65 &  \\
AlP         &   -0.99 &  -11.50 &    4.27 &   -0.79 &    3.86 &   -2.17 &    2.58 &  \\
AlAs        &   -1.00 &  -11.96 &    2.96 &   -0.83 &    3.12 &   -2.20 &    2.39 &  \\
AlSb        &   -0.78 &  -10.79 &    2.38 &   -0.93 &    2.10 &   -2.23 &    1.93 &  \\
GaN         &   -1.30 &  -16.41 &    3.11 &   -0.98 &    6.25 &   -2.71 &    4.88 &  -16.82 \\
wz          &   -1.33 &         &    3.48 &         &         &         &         &  -16.57 \\
GaP         &   -0.85 &  -12.51 &    2.63 &   -1.15 &    2.50 &   -2.74 &    2.39 &  -17.98 \\
GaAs        &   -0.78 &  -12.75 &    1.19 &   -1.16 &    1.68 &   -2.71 &    2.07 &  -18.30 \\
GaSb        &   -0.66 &  -11.57 &    0.62 &   -1.18 &    0.92 &   -2.63 &    1.37 &  -18.66 \\
InP         &   -0.76 &  -11.47 &    1.23 &   -1.01 &    2.12 &   -2.38 &    2.34 &  -16.48 \\
InAs        &   -0.86 &  -11.72 &    0.32 &   -0.91 &    1.65 &   -2.24 &    2.24 &  -16.47 \\
InSb        &   -0.68 &  -10.71 &    0.26 &   -1.02 &    1.03 &   -2.29 &    1.82 &  -16.87 \\
MgO         &   -2.42 &  -18.21 &    8.04 &   -0.73 &   11.42 &   -1.49 &   12.48 &         \\
ZnO         &   -2.17 &  -18.11 &    3.10 &   -0.81 &    8.20 &   -2.21 &    7.71 &   -6.68 \\
wz          &   -2.30 &         &    3.41 &         &         &         &         &   -6.47 \\
ZnS         &   -1.43 &  -12.82 &    3.66 &   -0.86 &    5.01 &   -2.21 &    4.92 &   -8.27 \\
ZnSe        &   -1.37 &  -13.13 &    2.64 &   -0.85 &    4.03 &   -2.14 &    4.34 &   -8.62 \\
ZnTe        &   -1.03 &  -11.75 &    2.35 &   -0.89 &    2.89 &   -2.16 &    3.32 &   -9.14 \\
CdS         &   -1.22 &  -12.16 &    2.30 &   -0.79 &    4.37 &   -1.94 &    4.77 &   -9.18 \\
CdSe        &   -1.16 &  -12.53 &    1.55 &   -0.76 &    3.52 &   -1.85 &    4.19 &   -9.39 \\
CdTe        &   -0.95 &  -11.17 &    1.62 &   -0.79 &    2.76 &   -1.88 &    3.46 &   -9.87 
\end{tabular}
\end{ruledtabular}
\end{table}

\begin{figure}
    \begin{center}
    \includegraphics[width=7cm,clip=true]{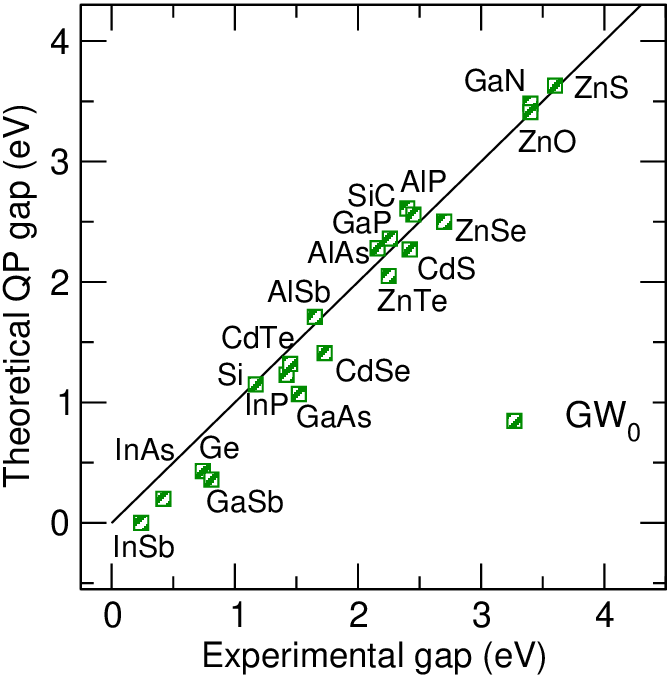}
    \end{center}
    \caption{ 
 Band gaps predicted using $GW_0$@LDA compared to experimental values. 
 The theoretical band gaps have been corrected for spin orbital coupling given in the 
 supplementary material  of Ref.~\onlinecite{grueneis2014}. Experimental data 
 are also collected in Ref.~\onlinecite{grueneis2014}.
    \label{fig:gap}}
\end{figure}

Fig. \ref{fig:gap} compares the calculated  $GW_0$@LDA tabulated in Tab.~\ref{tab:GW0}
with experimental values collected in Ref.~\onlinecite{grueneis2014}. 
We observe that the band gaps for
the ``metallic'' LDA-systems  (InSb, InAs, GaSb, and Ge) are noticeably too small. 
For $sp$ bonded systems the agreement is seemingly rather good,
although it should be noted that electron-phonon coupling can significantly reduce
the experimentally measured gap with the effect being about 0.4~eV for
C and 0.1~eV for Si.\cite{giustino2010,cannuccia2011,ponce2014}
We expect similar contributions of the order of 0.4~eV for
other oxides, nitrides, and carbides, and of the order of 0.1--0.2 for sulfides and materials containing aluminium. 
If these corrections are considered the comparison with experiment would
be less favorable. 
A final answer, however, requires one to perform
the actual calculations with the phononic contributions accounted for.

\section{Discussion and Conclusion}

One central result of the present work is a simple asymptotic estimate
of the self-energy contribution  
from high energy plane wave like orbitals, see. Eq.~(\ref{equ:exactGW}). 
Remarkably, this estimate allows to determine  the error in QP corrections
using only the total charge density and the charge density of
the QP state in question. 
The estimate is exceedingly  easy to calculate and can be readily 
implemented in any plane wave or all-electron code. 
A similar error estimate is obtained for the correlation energies, given by
Eq.~(\ref{eq_dmp2_rho2}).
That both errors are related is fairly obvious, since QP removal and addition energies
can be regarded as the energy difference between a system with $N$ and
$N-1$ and $N+1$ electrons, respectively. 
The important observation is that in both cases, plane wave like orbitals $G$ and $G'$, that are 
not relevant for groundstate calculations, contribute to the correlation energy,
since the density at a wave vector $G-G'$ is involved.
Due to this folding, even plane waves at very high kinetic 
energies contribute to the correlation energy. 
We have discussed briefly that the origin of this problem is related to the so 
called cusp condition. 
The addition of an electron at position $r$ induces a charge depletion around $r$, 
which is essentially described by the exchange-correlation hole $h^{\rm xc}(r,r')$.
The correlation contribution to the hole is related to the screening charge density, and
these quantities have a cusp for $r'\rightarrow r$ which is correctly accounted for only
when an infinitely large basis set is used.

The present correction is complementary to resolution of identity methods\cite{bruneval2008} or
effective energy denominator techniques.\cite{berger2012}
These techniques allow to obtain converged results for a specific plane wave
basis set using only a small number of orbitals. They, however,
do not allow to determine the error incurred by disregarding 
plane wave components in the basis set. 
Our present correction does exactly this. 
Specifically, it allows to estimate the residual error resulting 
from the neglect of high energy plane wave components in the response function.
As observed before (and rederived here) this error is proportional to the inverse
of the number of basis functions included in the response function.
The correction has been tested for the case of ZnO and seems to be quite promising,
although, further refinements will be required for $d$ states.

Alternatively, one can extrapolate to the infinite basis set limit
by performing a series of calculations with an increasing basis set
for the orbitals and the response function. According to the derived asymptotic behaviour,
the error should fall off like one over the number of basis
functions. This procedure has already been applied by Garcia and Marini 
in RPA calculations.\cite{garcia2007} 
To avoid any bias, this procedure was adopted in the present work 
to predict QP energies for 24 materials.

Unfortunately, the derivation of  Eq.~(\ref{equ:exactGW})
indicates that standard PAW potentials with 
{\it non} norm-conserving ``ultrasoft'' partial waves will not describe QP levels
accurately, since the norm is not correctly restored when high energy
plane wave like orbitals are involved.  We observe that the resulting error can be substantial.
For instance, for ZnO using US PAW potentials, 
the $d$ levels are almost 1~eV too high in energy compared to accurate reference calculations,
and the band gap error is about 400~meV.
Although ZnO is an extreme example, since the $d$ levels are rather shallow 
(7~eV below the Fermi level) and the violation of the norm is particularly 
large for $3d$ elements, we found  that similar errors are observed for all materials 
containing elements with $3d$, as well as $4d$ and $5d$ electrons. 
Even for AlAs and GaAs, the inclusion of the  As $3d$ levels changes the indirect gap by
up to 200~meV.
Eq.~(\ref{equ:exactGW}) suggests two strategies to reduce the error: 
(i) either more projectors  at higher energies need to be included, or alternatively 
(ii) the partial waves are made norm-conserving. 
Although, we have not discussed option
(i) in detail, we found that it is, in practice, difficult to increase the number
of projectors beyond three for a given quantum number $l$. 
Solution (ii), however, works  reliably and yields  PAW potentials
that are robust and accurate. Furthermore, the specific choice for the core/pseudization 
radius
seems to influence the QP energies only very little for NC PAW potentials, 
as demonstrated for ZnO (and observed for all materials considered here).
The disadvantage of NC PAW potentials is that they require about a
factor of 2 larger plane wave cut-off energies than standard ultrasoft PAW potentials. 
Although this makes the potentials impractical for
groundstate calculations, fairly large plane wave cut-offs are anyway
required to obtain converged QP energies, so that the increased cut-off
has, in practice, little consequences  for $GW$ calculations.

In this work, we calculated $G_0W_0$ and $GW_0$ band gaps for prototypical 
semiconductors and insulators using the newly constructed NC PAW potentials.
Overall, we observe what has been known for quite some time.
The QP gaps for medium gap materials are very accurate using
the $G_0W_0$ approximation, slightly outperformed by the $GW_0$ approximation.
Errors are even acceptable for ZnO, that has long been considered to
be problematic. Indeed the main issue with ZnO was insufficient basis set
convergence as already noted in Refs.~\onlinecite{shih2010} and~\onlinecite{friedrich2011}.
For small gap systems, however, our results are clearly unsatisfactory, and this concerns,
in particular, materials that show incorrect band  order at the $\Gamma$ point
in the local density approximation: in Ge, GaSb, InAs, InSb, and CdO,
the anion $p$ states are above the Fermi-level, whereas the
cation $s$ state is below the Fermi-level for LDA. 
This results in sizable errors in
the standard perturbative $GW$ approach.
Clearly better starting points than DFT orbitals and  DFT screening are required in these cases. 
Presently the most successful approaches are the self-consistent procedures
of Kotani and Schilfgaarde\cite{faleev2004,schilfgaarde2006}
or hybrid functionals if efficiency is important.\cite{fuchs2007,paier2008}
Moreover, we recently performed $GW$ calculations based on RPA optimized effective potential calculations
or hybrid functionals where the extrapolation procedures were applied as well.\cite{klimes2014OEP,grueneis2014}

We believe that our reported values are as accurate
as we can possibly make them within the limits of the $GW$@DFT approximation, and we hope that they can serve
as a benchmark for other $GW$ codes. 
It is in fact uttermost time for the ab initio community to establish 
such benchmarks to make $G_0W_0$ a truly validated tool (as DFT already is). 
After all, validation has made quantum chemistry approaches using
Gaussian type orbitals and DFT so successful. 
Unfortunately, $GW$ results with that kind of reference quality
are still  rare. Specifically, we are only aware of accurate calculations for Si and ZnO, 
for which we have made a comparison and found very good agreement.\cite{friedrich2006,friedrich2011}
We also want to emphasize that methods
employing compact basis sets, such as the linearized
muffin tin orbital methods and even the standard full potential linearized plane wave 
method will experience similar problems to the ones we experience with
non norm-conserving PAW potentials: to obtain highly accurate QP corrections and
many electron correlation energies, plane waves and local basis functions that
do not contribute to the groundstate orbitals need to
be included in the excited state calculations.
This fact was known in the quantum chemistry community for some time and certainly
imposes a computational challenge that needs to be overcome
when reference quality data are sought. 

\begin{acknowledgments}

This work was supported by the Austrian Science Fund (FWF)  within the SFB ViCoM (Grant F 41).
Supercomputing time on the Vienna Scientific cluster (VSC) is gratefully acknowledged.
\end{acknowledgments}

\bibliography{GW}

\begin{thebibliography}{46}
\expandafter\ifx\csname natexlab\endcsname\relax\def\natexlab#1{#1}\fi
\expandafter\ifx\csname bibnamefont\endcsname\relax
  \def\bibnamefont#1{#1}\fi
\expandafter\ifx\csname bibfnamefont\endcsname\relax
  \def\bibfnamefont#1{#1}\fi
\expandafter\ifx\csname citenamefont\endcsname\relax
  \def\citenamefont#1{#1}\fi
\expandafter\ifx\csname url\endcsname\relax
  \def\url#1{\texttt{#1}}\fi
\expandafter\ifx\csname urlprefix\endcsname\relax\def\urlprefix{URL }\fi
\providecommand{\bibinfo}[2]{#2}
\providecommand{\eprint}[2][]{\url{#2}}

\bibitem[{\citenamefont{Hedin}(1965)}]{hedin1965}
\bibinfo{author}{\bibfnamefont{L.}~\bibnamefont{Hedin}},
  \bibinfo{journal}{Phys. Rev.} \textbf{\bibinfo{volume}{139}},
  \bibinfo{pages}{A796} (\bibinfo{year}{1965}).

\bibitem[{\citenamefont{Hybertsen and Louie}(1986)}]{hybertsen1986}
\bibinfo{author}{\bibfnamefont{M.~S.} \bibnamefont{Hybertsen}}
  \bibnamefont{and} \bibinfo{author}{\bibfnamefont{S.~G.} \bibnamefont{Louie}},
  \bibinfo{journal}{Phys. Rev. B} \textbf{\bibinfo{volume}{34}},
  \bibinfo{pages}{5390} (\bibinfo{year}{1986}).

\bibitem[{\citenamefont{Lannoo et~al.}({1985})\citenamefont{Lannoo,
  Schl\"{u}ter, and Sham}}]{lannoo1985}
\bibinfo{author}{\bibfnamefont{M.}~\bibnamefont{Lannoo}},
  \bibinfo{author}{\bibfnamefont{M.}~\bibnamefont{Schl\"{u}ter}},
  \bibnamefont{and} \bibinfo{author}{\bibfnamefont{L.~J.} \bibnamefont{Sham}},
  \bibinfo{journal}{Phys. Rev. B} \textbf{\bibinfo{volume}{{32}}},
  \bibinfo{pages}{{3890}} (\bibinfo{year}{{1985}}).

\bibitem[{\citenamefont{Aryasetiawan and Gunnarsson}(1998)}]{aryasetiawan1998}
\bibinfo{author}{\bibfnamefont{F.}~\bibnamefont{Aryasetiawan}}
  \bibnamefont{and}
  \bibinfo{author}{\bibfnamefont{O.}~\bibnamefont{Gunnarsson}},
  \bibinfo{journal}{Rep. Prog. Phys.} \textbf{\bibinfo{volume}{61}},
  \bibinfo{pages}{237} (\bibinfo{year}{1998}).

\bibitem[{\citenamefont{Onida et~al.}(2002)\citenamefont{Onida, Reining, and
  Rubio}}]{onida2002}
\bibinfo{author}{\bibfnamefont{G.}~\bibnamefont{Onida}},
  \bibinfo{author}{\bibfnamefont{L.}~\bibnamefont{Reining}}, \bibnamefont{and}
  \bibinfo{author}{\bibfnamefont{A.}~\bibnamefont{Rubio}},
  \bibinfo{journal}{Rev. Mod. Phys.} \textbf{\bibinfo{volume}{74}},
  \bibinfo{pages}{601} (\bibinfo{year}{2002}).

\bibitem[{\citenamefont{Giuliani and Vignale}(2005)}]{giuliani2005}
\bibinfo{author}{\bibfnamefont{G.}~\bibnamefont{Giuliani}} \bibnamefont{and}
  \bibinfo{author}{\bibfnamefont{G.}~\bibnamefont{Vignale}},
  \emph{\bibinfo{title}{Quantum Theory of the Electron Liquid}}
  (\bibinfo{publisher}{CUP, Cambridge}, \bibinfo{year}{2005}).

\bibitem[{\citenamefont{Bechstedt et~al.}(2009)\citenamefont{Bechstedt, Fuchs,
  and Kresse}}]{bechstedt2009}
\bibinfo{author}{\bibfnamefont{F.}~\bibnamefont{Bechstedt}},
  \bibinfo{author}{\bibfnamefont{F.}~\bibnamefont{Fuchs}}, \bibnamefont{and}
  \bibinfo{author}{\bibfnamefont{G.}~\bibnamefont{Kresse}},
  \bibinfo{journal}{Phys. Stat. Sol. B} \textbf{\bibinfo{volume}{246}},
  \bibinfo{pages}{1877} (\bibinfo{year}{2009}).

\bibitem[{\citenamefont{Starke and Kresse}(2012)}]{starke2012}
\bibinfo{author}{\bibfnamefont{R.}~\bibnamefont{Starke}} \bibnamefont{and}
  \bibinfo{author}{\bibfnamefont{G.}~\bibnamefont{Kresse}},
  \bibinfo{journal}{Phys. Rev. B} \textbf{\bibinfo{volume}{85}},
  \bibinfo{pages}{075119} (\bibinfo{year}{2012}).

\bibitem[{\citenamefont{Shih et~al.}(2010)\citenamefont{Shih, Xue, Zhang,
  Cohen, and Louie}}]{shih2010}
\bibinfo{author}{\bibfnamefont{B.-C.} \bibnamefont{Shih}},
  \bibinfo{author}{\bibfnamefont{Y.}~\bibnamefont{Xue}},
  \bibinfo{author}{\bibfnamefont{P.}~\bibnamefont{Zhang}},
  \bibinfo{author}{\bibfnamefont{M.~L.} \bibnamefont{Cohen}}, \bibnamefont{and}
  \bibinfo{author}{\bibfnamefont{S.~G.} \bibnamefont{Louie}},
  \bibinfo{journal}{Phys. Rev. Lett.} \textbf{\bibinfo{volume}{105}},
  \bibinfo{pages}{146401} (\bibinfo{year}{2010}).

\bibitem[{\citenamefont{Stankovski et~al.}(2011)\citenamefont{Stankovski,
  Antonius, Waroquiers, Miglio, Dixit, Sankaran, Giantomassi, Gonze, C\^ot\'e,
  and Rignanese}}]{stankovski2011}
\bibinfo{author}{\bibfnamefont{M.}~\bibnamefont{Stankovski}},
  \bibinfo{author}{\bibfnamefont{G.}~\bibnamefont{Antonius}},
  \bibinfo{author}{\bibfnamefont{D.}~\bibnamefont{Waroquiers}},
  \bibinfo{author}{\bibfnamefont{A.}~\bibnamefont{Miglio}},
  \bibinfo{author}{\bibfnamefont{H.}~\bibnamefont{Dixit}},
  \bibinfo{author}{\bibfnamefont{K.}~\bibnamefont{Sankaran}},
  \bibinfo{author}{\bibfnamefont{M.}~\bibnamefont{Giantomassi}},
  \bibinfo{author}{\bibfnamefont{X.}~\bibnamefont{Gonze}},
  \bibinfo{author}{\bibfnamefont{M.}~\bibnamefont{C\^ot\'e}}, \bibnamefont{and}
  \bibinfo{author}{\bibfnamefont{G.-M.} \bibnamefont{Rignanese}},
  \bibinfo{journal}{Phys. Rev. B} \textbf{\bibinfo{volume}{84}},
  \bibinfo{pages}{241201} (\bibinfo{year}{2011}).

\bibitem[{\citenamefont{Kato}(1957)}]{kato1957}
\bibinfo{author}{\bibfnamefont{T.}~\bibnamefont{Kato}},
  \bibinfo{journal}{Commun. Pure Appl. Math.} \textbf{\bibinfo{volume}{10}},
  \bibinfo{pages}{151} (\bibinfo{year}{1957}).

\bibitem[{\citenamefont{Schwartz}(1962)}]{schwartz1962}
\bibinfo{author}{\bibfnamefont{C.}~\bibnamefont{Schwartz}},
  \bibinfo{journal}{Phys. Rev.} \textbf{\bibinfo{volume}{126}},
  \bibinfo{pages}{1015} (\bibinfo{year}{1962}).

\bibitem[{\citenamefont{Kutzelnigg and Morgan}(1992)}]{kutzelnigg1992}
\bibinfo{author}{\bibfnamefont{W.}~\bibnamefont{Kutzelnigg}} \bibnamefont{and}
  \bibinfo{author}{\bibfnamefont{J.~D.} \bibnamefont{Morgan}},
  \bibinfo{journal}{J. Chem. Phys.} \textbf{\bibinfo{volume}{96}},
  \bibinfo{pages}{4484} (\bibinfo{year}{1992}).

\bibitem[{\citenamefont{Kendall et~al.}(1992)\citenamefont{Kendall, Dunning,
  and Harrison}}]{kendall1992}
\bibinfo{author}{\bibfnamefont{R.~A.} \bibnamefont{Kendall}},
  \bibinfo{author}{\bibfnamefont{T.~H.} \bibnamefont{Dunning}},
  \bibnamefont{and} \bibinfo{author}{\bibfnamefont{R.~J.}
  \bibnamefont{Harrison}}, \bibinfo{journal}{J. Chem. Phys.}
  \textbf{\bibinfo{volume}{96}}, \bibinfo{pages}{6796} (\bibinfo{year}{1992}).

\bibitem[{\citenamefont{Halkier et~al.}(1998)\citenamefont{Halkier, Helgaker,
  J\/{o}rgensen, Klopper, Koch, Olsen, and Wilson}}]{halkier1998}
\bibinfo{author}{\bibfnamefont{A.}~\bibnamefont{Halkier}},
  \bibinfo{author}{\bibfnamefont{T.}~\bibnamefont{Helgaker}},
  \bibinfo{author}{\bibfnamefont{T.}~\bibnamefont{J\/{o}rgensen}},
  \bibinfo{author}{\bibfnamefont{W.}~\bibnamefont{Klopper}},
  \bibinfo{author}{\bibfnamefont{H.}~\bibnamefont{Koch}},
  \bibinfo{author}{\bibfnamefont{J.}~\bibnamefont{Olsen}}, \bibnamefont{and}
  \bibinfo{author}{\bibfnamefont{A.~K.} \bibnamefont{Wilson}},
  \bibinfo{journal}{Chem. Phys. Lett.} \textbf{\bibinfo{volume}{286}},
  \bibinfo{pages}{243} (\bibinfo{year}{1998}).

\bibitem[{\citenamefont{Harl and Kresse}(2008)}]{harl2008}
\bibinfo{author}{\bibfnamefont{J.}~\bibnamefont{Harl}} \bibnamefont{and}
  \bibinfo{author}{\bibfnamefont{G.}~\bibnamefont{Kresse}},
  \bibinfo{journal}{Phys. Rev. B} \textbf{\bibinfo{volume}{77}},
  \bibinfo{pages}{045136} (\bibinfo{year}{2008}).

\bibitem[{\citenamefont{Shepherd et~al.}(2012)\citenamefont{Shepherd,
  Gr\"uneis, Booth, Kresse, and Alavi}}]{shepherd2012}
\bibinfo{author}{\bibfnamefont{J.~J.} \bibnamefont{Shepherd}},
  \bibinfo{author}{\bibfnamefont{A.}~\bibnamefont{Gr\"uneis}},
  \bibinfo{author}{\bibfnamefont{G.~H.} \bibnamefont{Booth}},
  \bibinfo{author}{\bibfnamefont{G.}~\bibnamefont{Kresse}}, \bibnamefont{and}
  \bibinfo{author}{\bibfnamefont{A.}~\bibnamefont{Alavi}},
  \bibinfo{journal}{Phys. Rev. B} \textbf{\bibinfo{volume}{86}},
  \bibinfo{pages}{035111} (\bibinfo{year}{2012}).

\bibitem[{\citenamefont{Gulans}()}]{gulans_unp}
\bibinfo{author}{\bibfnamefont{A.}~\bibnamefont{Gulans}},
  \bibinfo{note}{unpublished}.

\bibitem[{\citenamefont{Bj\"{o}rkman et~al.}(2012)\citenamefont{Bj\"{o}rkman,
  Gulans, Krasheninnikov, and Nieminen}}]{bjorkman2012}
\bibinfo{author}{\bibfnamefont{T.}~\bibnamefont{Bj\"{o}rkman}},
  \bibinfo{author}{\bibfnamefont{A.}~\bibnamefont{Gulans}},
  \bibinfo{author}{\bibfnamefont{A.~V.} \bibnamefont{Krasheninnikov}},
  \bibnamefont{and} \bibinfo{author}{\bibfnamefont{R.~M.}
  \bibnamefont{Nieminen}}, \bibinfo{journal}{Phys. Rev. Lett.}
  \textbf{\bibinfo{volume}{108}}, \bibinfo{pages}{{235502}}
  (\bibinfo{year}{2012}).

\bibitem[{\citenamefont{Kang and Hybertsen}(2010)}]{kang2010}
\bibinfo{author}{\bibfnamefont{W.}~\bibnamefont{Kang}} \bibnamefont{and}
  \bibinfo{author}{\bibfnamefont{M.~S.} \bibnamefont{Hybertsen}},
  \bibinfo{journal}{Phys. Rev. B} \textbf{\bibinfo{volume}{82}},
  \bibinfo{pages}{195108} (\bibinfo{year}{2010}).

\bibitem[{\citenamefont{Friedrich et~al.}(2011)\citenamefont{Friedrich,
  M\"uller, and Bl\"ugel}}]{friedrich2011}
\bibinfo{author}{\bibfnamefont{C.}~\bibnamefont{Friedrich}},
  \bibinfo{author}{\bibfnamefont{M.~C.} \bibnamefont{M\"uller}},
  \bibnamefont{and} \bibinfo{author}{\bibfnamefont{S.}~\bibnamefont{Bl\"ugel}},
  \bibinfo{journal}{Phys. Rev. B} \textbf{\bibinfo{volume}{83}},
  \bibinfo{pages}{081101} (\bibinfo{year}{2011}).

\bibitem[{\citenamefont{Perdew et~al.}(1996)\citenamefont{Perdew, Burke, and
  Ernzerhof}}]{perdew1996}
\bibinfo{author}{\bibfnamefont{J.~P.} \bibnamefont{Perdew}},
  \bibinfo{author}{\bibfnamefont{K.}~\bibnamefont{Burke}}, \bibnamefont{and}
  \bibinfo{author}{\bibfnamefont{M.}~\bibnamefont{Ernzerhof}},
  \bibinfo{journal}{Phys. Rev. Lett.} \textbf{\bibinfo{volume}{77}},
  \bibinfo{pages}{3865} (\bibinfo{year}{1996}), \bibinfo{note}{$ibid$, {\bf
  78}, 1396 (1997)}.

\bibitem[{\citenamefont{Marsman et~al.}(2009)\citenamefont{Marsman, Gr\"uneis,
  Paier, and Kresse}}]{marsman2009}
\bibinfo{author}{\bibfnamefont{M.}~\bibnamefont{Marsman}},
  \bibinfo{author}{\bibfnamefont{A.}~\bibnamefont{Gr\"uneis}},
  \bibinfo{author}{\bibfnamefont{J.}~\bibnamefont{Paier}}, \bibnamefont{and}
  \bibinfo{author}{\bibfnamefont{G.}~\bibnamefont{Kresse}},
  \bibinfo{journal}{J. Chem. Phys.} \textbf{\bibinfo{volume}{130}},
  \bibinfo{eid}{184103} (\bibinfo{year}{2009}).

\bibitem[{\citenamefont{Gr\"{u}neis et~al.}(2013)\citenamefont{Gr\"{u}neis,
  Shepherd, Alavi, Tew, and Booth}}]{gruneis2013}
\bibinfo{author}{\bibfnamefont{A.}~\bibnamefont{Gr\"{u}neis}},
  \bibinfo{author}{\bibfnamefont{J.~J.} \bibnamefont{Shepherd}},
  \bibinfo{author}{\bibfnamefont{A.}~\bibnamefont{Alavi}},
  \bibinfo{author}{\bibfnamefont{D.~P.} \bibnamefont{Tew}}, \bibnamefont{and}
  \bibinfo{author}{\bibfnamefont{G.~H.} \bibnamefont{Booth}},
  \bibinfo{journal}{J. Chem. Phys.} \textbf{\bibinfo{volume}{139}},
  \bibinfo{pages}{084112} (\bibinfo{year}{2013}).

\bibitem[{\citenamefont{Bl\"ochl}(1994)}]{bloechl:94}
\bibinfo{author}{\bibfnamefont{P.}~\bibnamefont{Bl\"ochl}},
  \bibinfo{journal}{Phys. Rev. B} \textbf{\bibinfo{volume}{50}},
  \bibinfo{pages}{17953} (\bibinfo{year}{1994}).

\bibitem[{\citenamefont{Gr\"{u}neis et~al.}(2010)\citenamefont{Gr\"{u}neis,
  Marsman, and Kresse}}]{gruneis2010}
\bibinfo{author}{\bibfnamefont{A.}~\bibnamefont{Gr\"{u}neis}},
  \bibinfo{author}{\bibfnamefont{M.}~\bibnamefont{Marsman}}, \bibnamefont{and}
  \bibinfo{author}{\bibfnamefont{G.}~\bibnamefont{Kresse}},
  \bibinfo{journal}{J. Chem. Phys.} \textbf{\bibinfo{volume}{133}},
  \bibinfo{eid}{074107} (\bibinfo{year}{2010}).

\bibitem[{\citenamefont{Deslippe et~al.}(2013)\citenamefont{Deslippe,
  Samsonidze, Jain, Cohen, and Louie}}]{deslippe2013}
\bibinfo{author}{\bibfnamefont{J.}~\bibnamefont{Deslippe}},
  \bibinfo{author}{\bibfnamefont{G.}~\bibnamefont{Samsonidze}},
  \bibinfo{author}{\bibfnamefont{M.}~\bibnamefont{Jain}},
  \bibinfo{author}{\bibfnamefont{M.~L.} \bibnamefont{Cohen}}, \bibnamefont{and}
  \bibinfo{author}{\bibfnamefont{S.~G.} \bibnamefont{Louie}},
  \bibinfo{journal}{Phys. Rev. B} \textbf{\bibinfo{volume}{87}},
  \bibinfo{pages}{165124} (\bibinfo{year}{2013}).

\bibitem[{\citenamefont{Gr\"uneis et~al.}(2014)\citenamefont{Gr\"uneis, Kresse,
  Hinuma, and Oba}}]{grueneis2014}
\bibinfo{author}{\bibfnamefont{A.}~\bibnamefont{Gr\"uneis}},
  \bibinfo{author}{\bibfnamefont{G.}~\bibnamefont{Kresse}},
  \bibinfo{author}{\bibfnamefont{Y.}~\bibnamefont{Hinuma}}, \bibnamefont{and}
  \bibinfo{author}{\bibfnamefont{F.}~\bibnamefont{Oba}},
  \bibinfo{journal}{Phys. Rev. Lett.} \textbf{\bibinfo{volume}{112}},
  \bibinfo{pages}{{096401}} (\bibinfo{year}{2014}).

\bibitem[{\citenamefont{Shishkin and Kresse}(2006)}]{shishkin2006}
\bibinfo{author}{\bibfnamefont{M.}~\bibnamefont{Shishkin}} \bibnamefont{and}
  \bibinfo{author}{\bibfnamefont{G.}~\bibnamefont{Kresse}},
  \bibinfo{journal}{Phys. Rev. B} \textbf{\bibinfo{volume}{74}},
  \bibinfo{pages}{035101} (\bibinfo{year}{2006}).

\bibitem[{\citenamefont{Shishkin and Kresse}(2007)}]{shishkin2007}
\bibinfo{author}{\bibfnamefont{M.}~\bibnamefont{Shishkin}} \bibnamefont{and}
  \bibinfo{author}{\bibfnamefont{G.}~\bibnamefont{Kresse}},
  \bibinfo{journal}{Phys. Rev. B} \textbf{\bibinfo{volume}{75}},
  \bibinfo{pages}{235102} (\bibinfo{year}{2007}).

\bibitem[{\citenamefont{Kresse and Joubert}(1999)}]{kresse99}
\bibinfo{author}{\bibfnamefont{G.}~\bibnamefont{Kresse}} \bibnamefont{and}
  \bibinfo{author}{\bibfnamefont{D.}~\bibnamefont{Joubert}},
  \bibinfo{journal}{Phys. Rev. B} \textbf{\bibinfo{volume}{59}},
  \bibinfo{pages}{1758} (\bibinfo{year}{1999}).

\bibitem[{\citenamefont{Klime\v{s} and Kresse}(2014)}]{klimes2014OEP}
\bibinfo{author}{\bibfnamefont{J.}~\bibnamefont{Klime\v{s}}} \bibnamefont{and}
  \bibinfo{author}{\bibfnamefont{G.}~\bibnamefont{Kresse}},
  \bibinfo{journal}{J. Chem. Phys.} \textbf{\bibinfo{volume}{104}},
  \bibinfo{pages}{054516} (\bibinfo{year}{2014}).

\bibitem[{\citenamefont{Bruneval and Gonze}({2008})}]{bruneval2008}
\bibinfo{author}{\bibfnamefont{F.}~\bibnamefont{Bruneval}} \bibnamefont{and}
  \bibinfo{author}{\bibfnamefont{X.}~\bibnamefont{Gonze}},
  \bibinfo{journal}{Phys. Rev. B} \textbf{\bibinfo{volume}{{78}}},
  \bibinfo{pages}{{085125}} (\bibinfo{year}{{2008}}).

\bibitem[{\citenamefont{Berger et~al.}({2012})\citenamefont{Berger, Reining,
  and Sottile}}]{berger2012}
\bibinfo{author}{\bibfnamefont{J.~A.} \bibnamefont{Berger}},
  \bibinfo{author}{\bibfnamefont{L.}~\bibnamefont{Reining}}, \bibnamefont{and}
  \bibinfo{author}{\bibfnamefont{F.}~\bibnamefont{Sottile}},
  \bibinfo{journal}{Phys. Rev. B} \textbf{\bibinfo{volume}{{85}}},
  \bibinfo{pages}{{085126}} (\bibinfo{year}{{2012}}).

\bibitem[{\citenamefont{Friedrich et~al.}(2006)\citenamefont{Friedrich,
  Schindlmayr, Bl\"ugel, and Kotani}}]{friedrich2006}
\bibinfo{author}{\bibfnamefont{C.}~\bibnamefont{Friedrich}},
  \bibinfo{author}{\bibfnamefont{A.}~\bibnamefont{Schindlmayr}},
  \bibinfo{author}{\bibfnamefont{S.}~\bibnamefont{Bl\"ugel}}, \bibnamefont{and}
  \bibinfo{author}{\bibfnamefont{T.}~\bibnamefont{Kotani}},
  \bibinfo{journal}{Phys. Rev. B} \textbf{\bibinfo{volume}{74}},
  \bibinfo{pages}{045104} (\bibinfo{year}{2006}).

\bibitem[{\citenamefont{Li et~al.}(2012)\citenamefont{Li, G\/omez-Abal, Jiang,
  Ambrosch-Draxl, and Scheffler}}]{li2012NJP}
\bibinfo{author}{\bibfnamefont{X.-Z.} \bibnamefont{Li}},
  \bibinfo{author}{\bibfnamefont{R.}~\bibnamefont{G\/omez-Abal}},
  \bibinfo{author}{\bibfnamefont{H.}~\bibnamefont{Jiang}},
  \bibinfo{author}{\bibfnamefont{C.}~\bibnamefont{Ambrosch-Draxl}},
  \bibnamefont{and}
  \bibinfo{author}{\bibfnamefont{M.}~\bibnamefont{Scheffler}},
  \bibinfo{journal}{New J. Phys.} \textbf{\bibinfo{volume}{14}},
  \bibinfo{pages}{023006} (\bibinfo{year}{2012}).

\bibitem[{\citenamefont{G\'omez-Abal et~al.}(2008)\citenamefont{G\'omez-Abal,
  Li, Scheffler, and Ambrosch-Draxl}}]{abal2008}
\bibinfo{author}{\bibfnamefont{R.}~\bibnamefont{G\'omez-Abal}},
  \bibinfo{author}{\bibfnamefont{X.}~\bibnamefont{Li}},
  \bibinfo{author}{\bibfnamefont{M.}~\bibnamefont{Scheffler}},
  \bibnamefont{and}
  \bibinfo{author}{\bibfnamefont{C.}~\bibnamefont{Ambrosch-Draxl}},
  \bibinfo{journal}{Phys. Rev. Lett.} \textbf{\bibinfo{volume}{101}},
  \bibinfo{pages}{106404} (\bibinfo{year}{2008}).

\bibitem[{\citenamefont{Friedrich et~al.}(2010)\citenamefont{Friedrich,
  Bl\"ugel, and Schindlmayr}}]{friedrich2010}
\bibinfo{author}{\bibfnamefont{C.}~\bibnamefont{Friedrich}},
  \bibinfo{author}{\bibfnamefont{S.}~\bibnamefont{Bl\"ugel}}, \bibnamefont{and}
  \bibinfo{author}{\bibfnamefont{A.}~\bibnamefont{Schindlmayr}},
  \bibinfo{journal}{Phys. Rev. B} \textbf{\bibinfo{volume}{81}},
  \bibinfo{pages}{125102} (\bibinfo{year}{2010}).

\bibitem[{\citenamefont{Giustino et~al.}({2010})\citenamefont{Giustino, Louie,
  and Cohen}}]{giustino2010}
\bibinfo{author}{\bibfnamefont{F.}~\bibnamefont{Giustino}},
  \bibinfo{author}{\bibfnamefont{S.~G.} \bibnamefont{Louie}}, \bibnamefont{and}
  \bibinfo{author}{\bibfnamefont{M.~L.} \bibnamefont{Cohen}},
  \bibinfo{journal}{Phys. Rev. Lett.} \textbf{\bibinfo{volume}{{105}}},
  \bibinfo{pages}{{265501}} (\bibinfo{year}{{2010}}).

\bibitem[{\citenamefont{Cannuccia and Marini}({2011})}]{cannuccia2011}
\bibinfo{author}{\bibfnamefont{E.}~\bibnamefont{Cannuccia}} \bibnamefont{and}
  \bibinfo{author}{\bibfnamefont{A.}~\bibnamefont{Marini}},
  \bibinfo{journal}{Phys. Rev. Lett.} \textbf{\bibinfo{volume}{{107}}},
  \bibinfo{pages}{{255501}} (\bibinfo{year}{{2011}}).

\bibitem[{\citenamefont{Ponc\'{e} et~al.}(2014)\citenamefont{Ponc\'{e},
  Antonius, Boulanger, Cannuccia, Marini, C\^{o}t\'{e}, and Gonze}}]{ponce2014}
\bibinfo{author}{\bibfnamefont{S.}~\bibnamefont{Ponc\'{e}}},
  \bibinfo{author}{\bibfnamefont{G.}~\bibnamefont{Antonius}},
  \bibinfo{author}{\bibfnamefont{P.}~\bibnamefont{Boulanger}},
  \bibinfo{author}{\bibfnamefont{E.}~\bibnamefont{Cannuccia}},
  \bibinfo{author}{\bibfnamefont{A.}~\bibnamefont{Marini}},
  \bibinfo{author}{\bibfnamefont{M.}~\bibnamefont{C\^{o}t\'{e}}},
  \bibnamefont{and} \bibinfo{author}{\bibfnamefont{X.}~\bibnamefont{Gonze}},
  \bibinfo{journal}{Comp. Mat. Sci.} \textbf{\bibinfo{volume}{83}},
  \bibinfo{pages}{{341}} (\bibinfo{year}{2014}).

\bibitem[{\citenamefont{Garc{\'\i}a-Gonz\'{a}lez
  et~al.}(2007)\citenamefont{Garc{\'\i}a-Gonz\'{a}lez, Fern\'{a}ndez, Marini,
  and Rubio}}]{garcia2007}
\bibinfo{author}{\bibfnamefont{P.}~\bibnamefont{Garc{\'\i}a-Gonz\'{a}lez}},
  \bibinfo{author}{\bibfnamefont{J.~J.} \bibnamefont{Fern\'{a}ndez}},
  \bibinfo{author}{\bibfnamefont{A.}~\bibnamefont{Marini}}, \bibnamefont{and}
  \bibinfo{author}{\bibfnamefont{A.}~\bibnamefont{Rubio}}, \bibinfo{journal}{J.
  Phys. Chem. A} \textbf{\bibinfo{volume}{111}}, \bibinfo{pages}{12458}
  (\bibinfo{year}{2007}).

\bibitem[{\citenamefont{Faleev et~al.}(2004)\citenamefont{Faleev, van
  Schilfgaarde, and Kotani}}]{faleev2004}
\bibinfo{author}{\bibfnamefont{S.~V.} \bibnamefont{Faleev}},
  \bibinfo{author}{\bibfnamefont{M.}~\bibnamefont{van Schilfgaarde}},
  \bibnamefont{and} \bibinfo{author}{\bibfnamefont{T.}~\bibnamefont{Kotani}},
  \bibinfo{journal}{Phys. Rev. Lett.} \textbf{\bibinfo{volume}{93}},
  \bibinfo{pages}{126406} (\bibinfo{year}{2004}).

\bibitem[{\citenamefont{van Schilfgaarde et~al.}(2006)\citenamefont{van
  Schilfgaarde, Kotani, and Faleev}}]{schilfgaarde2006}
\bibinfo{author}{\bibfnamefont{M.}~\bibnamefont{van Schilfgaarde}},
  \bibinfo{author}{\bibfnamefont{T.}~\bibnamefont{Kotani}}, \bibnamefont{and}
  \bibinfo{author}{\bibfnamefont{S.}~\bibnamefont{Faleev}},
  \bibinfo{journal}{Phys. Rev. Lett.} \textbf{\bibinfo{volume}{96}},
  \bibinfo{pages}{226402} (\bibinfo{year}{2006}).

\bibitem[{\citenamefont{Fuchs et~al.}(2007)\citenamefont{Fuchs, Furthm\"uller,
  Bechstedt, Shishkin, and Kresse}}]{fuchs2007}
\bibinfo{author}{\bibfnamefont{F.}~\bibnamefont{Fuchs}},
  \bibinfo{author}{\bibfnamefont{J.}~\bibnamefont{Furthm\"uller}},
  \bibinfo{author}{\bibfnamefont{F.}~\bibnamefont{Bechstedt}},
  \bibinfo{author}{\bibfnamefont{M.}~\bibnamefont{Shishkin}}, \bibnamefont{and}
  \bibinfo{author}{\bibfnamefont{G.}~\bibnamefont{Kresse}},
  \bibinfo{journal}{Phys. Rev. B} \textbf{\bibinfo{volume}{76}},
  \bibinfo{pages}{115109} (\bibinfo{year}{2007}).

\bibitem[{\citenamefont{Paier et~al.}(2008)\citenamefont{Paier, Marsman, and
  Kresse}}]{paier2008}
\bibinfo{author}{\bibfnamefont{J.}~\bibnamefont{Paier}},
  \bibinfo{author}{\bibfnamefont{M.}~\bibnamefont{Marsman}}, \bibnamefont{and}
  \bibinfo{author}{\bibfnamefont{G.}~\bibnamefont{Kresse}},
  \bibinfo{journal}{Phys. Rev. B} \textbf{\bibinfo{volume}{78}},
  \bibinfo{pages}{121201} (\bibinfo{year}{2008}).

\end{thebibliography}

\end{document}